# Astronomy's Greatest Hits:

# The 100 most Cited Papers in Each Year of the First Decade of the 21$^{st}$ Century (2000 – 2009)

## Jay A. Frogel

Association of Universities for Research in Astronomy

1212 New York Avenue, NW, Suite 450

Washington, DC  20005

jfrogel@AURA-astronomy.org

# ABSTRACT

The first decade of the 21st Century and the last few years of the 20th have been transformative for ground and space based observational astronomy due to new observing facilities, access to digital archives, and growth in use of the Internet for communication and dissemination of information and for access to the archives. How have these three factors affected the characteristics and content of papers published in refereed astronomical journals as well as the journals themselves? In this and subsequent papers I will propose answers to this question.

The analysis in this the first paper of a series is based on an examination of the 100 most cited papers in astronomy and astrophysics for each year between 2000 and 2009, inclusive, and supplemental data from 1995 and 1990. The main findings of this analysis are:

Over the ten-year period the *total* number of authors of the top 100 articles per year has more than tripled. This increase is seen most strongly in papers with more than 6 authors. The number of *unique* authors in any given year has more than doubled. The yearly number of papers with 5 or fewer authors has declined over the same time period.

Averaged over the ten-year period the normalized number of authors *per paper* increases steadily with citation rank – the most highly cited papers tend to have the largest number of authors and visa versa. This increase is especially notable for papers ranked 1 through 20 in terms of number of citations *and* number of authors.

The *distribution* of normalized citation counts versus ranking is remarkably constant from year to year and, except for the top ranked half dozen or so papers in each year, is very closely approximated by a power law. Nearly all of the papers that show the most divergence from the power law fit – all in the sense of having a high number of citations – are based on the results of large observational surveys.

Amongst the top 100 papers there is a small but significant correlation of paper length with citation rank. More striking, though, is that the average page length of the top 100 papers is one and a half times that for astronomy papers in general.

For every year from 2000 to 2008 the same 5 journals account for 80 to 85% of the *total* citations for each year from all of the journals in the category "Astronomy and Astrophysics" by ISI's Journal Citation Reports. These numbers do not include *Nature* or *Science*. Averaged over the 10 year time period studied in this paper, these same 5 journals account for 77% of the 1000 most cited papers, slightly less than the journals' fractional contribution to the *total* number of articles published by all journals. The 5 journals are *A&A, AJ, ApJ, ApJS,* and *MNRAS*.

Two samples of the top 100 cited papers, both for the six years 2001 to 2006 but compiled two and a half years apart, show that a significant number of articles originally ranked in the top 100 for the year, drop out and are replaced by other articles as time passes. Most of the drop-outs address topics in extra-galactic astronomy; their replacements for the most part deal with non-extra-galactic topics.

Finally, some additional findings are noted that relate to the entire ensemble of astronomical journals published during the Century's first decade.

Various indicators of internet access to astronomical web sites such as data archives and journal repositories show increases of between factors of three and ten or more I propose that there are close complementarities between the communication capabilities that internet usage enables and the strong growth in numbers of authors of the most highly cited papers. Subsequent papers will examine this and other interpretations of the analysis presented here in detail.

# 1. INTRODUCTION

The first decade of the 21$^{st}$ Century and the last few years of the 20$^{th}$ Century have been transformative for observational astronomy. Three important reasons are new facilities, the digital archives that have resulted from them - as well as from older facilities - and the growth in use of the Internet for communication and dissemination of information and for enabling easy access to the archives. Intimately linked to all three of these is the steady increase in computing power as illustrated by the continuing validity of Moore's law (Moore, 1965).

First, consider the new facilities. On the ground essentially all of the existing 6-meter and larger telescopes have come into regular operation. In space the wide range of new missions includes Mars orbiters and landers, the Cassini mission to Saturn, Chandra, FUSE, XMM-Newton, WMAP, RHESSI, GALEX, Spitzer, Swift, Deep Impact, MESSENGER, Fermi, Kepler, Herschel, Planck, and WISE as well as Hubble Servicing Missions, and several high altitude ground and balloon-borne CMB telescopes. Some of these new capabilities have only just begun operations so they have not yet begun to produce a significant number of publications.

Second, enormous and readily accessible digital databases have resulted from major new observational surveys at all wavelengths from space and from the ground. Examples of ground based surveys include the SDSS, 2dF, 2MASS, Auger, HESS, and VERITAS. Some of these, while not necessarily producing large data sets, do result in highly cited papers. WMAP, Planck, WISE, and Fermi are new space missions with all sky surveys as their primary goal. And, as from the ground, targeted surveys on small areas have produced rich databases that are the subject of intensive study and have resulted in many highly cited publications. These include the Hubble Deep and Ultra-Deep Fields, GOODS, the several supernova surveys, and the search for extra-solar planetary systems.

Third, is the phenomenal increase in internet usage. The internet enables electronic access to nearly the entire body of astronomical literature (eg. Kurtz, et al., 2000 and Henneken et al., 2009), to most large repositories of data, and it permits essentially instantaneous communication across all geographical boundaries (cf. Abt 2000b). The internet also gives access to many tools for manipulating data and mining the astronomical archives.

What effects have these new facilities, data archives, and means of information exchange had on astronomical publications? In this paper, the first of a series, I will identify trends and patterns in astronomical publications that will be the basis for gauging these effects. This work is based on astronomy's "greatest hits", the 100 most cited papers in astronomy for each year of the last decade (2000 to 2009), 1,000 papers, by the top 100 papers for 1990 and 1995.

This paper concentrates on the inter-relationship between properties that characterize these top-cited papers and their dependence on time. Among the properties that I examine are the journal in which they are published, the number of authors on a paper, a paper's citation record, and page length. For the most part I will postpone interpretation of the findings to a subsequent paper which will discuss the scientific content of the papers, how, for observational papers, the data were acquired, and what role data archives played in the research.

Abt (2000b) gave several interesting answers to the question "what can we learn from publication studies", and pointed to a number of ways in which such studies can be put to practical use. The analysis in this paper will provide additional answers to his question that I hope will be of equal interest.

## 2. SOURCES FOR THE DATA AND THE ANALYSIS PROCEDURE

A simple, commonly used quantitative measure of the impact of a published account of a piece of research is the number of citations to the publication. The $h$-index (Hirsch 2005) is an often used statistic that attempts to quantify both the scientific output and impact of individuals and institutions. Recent examples of citation studies (as well as studies of "citation studies") may be found in the work of a number of authors; for example Abt (2000a, 2000b, 2003, 2006a, b, 2007), Meho (2007), Madrid and Macchetto (2004), Trimble and Ceja (2007, 2008, 2010), Van Noorden (2010), Kinney (2007, 2008), Kurtz, et al. (2005), Stanek (2008, 2009), White (2007), and Molinari and Molinari (2008). The goals of these and similar papers have included not only the evaluation of the impact of individuals, institutions, and telescopes, but also a search for broad trends in astronomical publications.

The quantitative analysis in the present paper is based on two overlapping samples of highly cited papers in astronomy and astrophysics, nearly all of which are in peer reviewed journals, but about 5% are in journals devoted to review articles. The first sample was compiled in mid-2007 and consists of the 100 most cited papers in astronomy and astrophysics for 2001 to 2006 inclusive. For brevity I will refer to this sample as the "six-100" sample. Data for the second sample – the one on which most of this paper is based - were assembled in early 2010 and consist of the 100 most cited papers in astronomy and astrophysics for each year in the past decade 2000 – 2009. This sample will be referred to as the "ten-100" sample or, simply, "the sample". A full description of how papers in the six-100 and ten-100 samples were selected is in Section 3. A comparison of citation counts from ISI and ADS is in Appendix A.

I will examine a number of statistics that characterize the papers in the two samples such as their number of authors and citations, their page length, and where they were published. Sections 3 through 6 will present the statistical analysis and the findings; Section 7 will summarize the findings and suggest some interpretations, and identify areas worth further study.

While citation counting is not the only way to gauge a paper's impact, it is a simple number to extract from publication data bases and conceptually provides a straightforward way of comparing papers. Abt (2000c) has in fact shown that "important papers almost invariably produce many more citations than others, and citation counts are good measures of importance or usefulness." An obvious shortcoming is that citation counts are time dependent. Proposals to minimize the impact of the "age" of a paper include the "$h$-index" and the "impact factor" (Hirsch 2005; Bordons, Fernández, and Gomez 2002, but see Van Noorden 2010).

As an alternative to the absolute number of citations and to the $h$-index, I have chosen to use the simple, non-parametric statistic of the *ranking* of a paper in terms of citation counts relative to other refereed papers published in the same year. Thus a paper ranked "1" in its sample for a year is the most cited in that sample *as of the date that the data were compiled*; larger ranking numbers mean smaller number of citations. It turns out that this statistic can also be time dependent, strongly so for the first few years immediately after publication. As will be seen in section 3.1.4 an examination of this time dependency yields some interesting results as well. Trimble and Ceja (2010) argue that the growth rate for citations to a paper can be strongly subject dependent. Furthermore, Abt (1996, 1998) has shown that the "half lives" of papers vary considerably. The results of the comparison of the ten-100 and six-100 samples in Section 3.1.4 probably reflect Trimble and Ceja's and Abt's findings. Nonetheless, rank has the advantage of minimizing the effects of outliers and of giving an overall picture of the influence of a wide range of papers. For these same reasons, I will also use a ranking statistic to investigate other parameters; again a "1" ranking will mean, for example, the largest number of authors or longest paper, etc. in the group of 100 for a particular year.

Finally, there is the issue of self-citations. Trimble (1986) found that "About 15% of all citations in astronomical papers published during January 1983 were self-citations, in the sense that the cited and citing papers had at least one author in common."  She found that this percentage varied little with parameters such as journal, country, topic, etc.  Based on her study, Abt (2000b) concluded that, although self-citations make up a non-negligible fraction of all references, they do not, on average, distort citation statistics significantly.  However, as this paper will show, the average number of authors per paper for the top 100 has increased by more than a factor of three over the past decade.  This, together with the increasing prevalence of papers with a large number (a dozen or more) of authors, probably means that the incidence of self-citations has gone up considerably.  Unfortunately, trying to correct for this effect is beyond the scope of this paper (or the time available to the author!).  Thus no corrections will be made for self-citations in the analysis which follows.

# 3.    SOURCES FOR THE DATA AND THE SELECTION OF JOURNALS AND PAPERS

## 3.1.    Selection of Journals and Extraction of the 100 Most Cited Articles

### 3.1.1.    Selection of Journals

In order to identify the refereed journals that publish papers on astronomy and astrophysics I used the Journal Citation Reports (JCR) on ISI's Web of Science to determine which journals account for the top ~95% of all citations for each year.  This was done twice for the two samples but separated in time by about two years. I selected the category "Astronomy & Astrophysics" on the JCR web page.  There were 34 such journals in 2000 and 43 in 2008.[1]  For each year I rank ordered the journals by total number of citations to articles published in that journal between the year of publication and late 2007 (for six-100) or early 2010 (for ten-100) – the times when I assembled the two data sets.  The rank ordering by total citation numbers of the journals year to year is quite consistent.  Typically, those that comprise the bottom half of the list account for 3% or less of the total number of citations for that year.  At the top, the same 5 journals consistently account for between 80 and 85% of the citations in any given year.  These 5 are the *ApJ, ApJS, A&A, AJ,* and *MNRAS*.  Wanting to err on the side of inclusiveness I selected those journals that together account for ~95% of the total citations to all journals in JCR's Astronomy and Astrophysics category in any given year.

After selecting the journals with 95% of the citations in a year, I sorted the remaining ones by "impact factor" (as defined by JCR)[2] and added to the initial list those that on average had impact factors greater than three.  This was to ensure that the study did not exclude small journals with only a few highly cited papers.  As a result, one journal was added to the list: *Astro-particle Physics*.  (Discussion of "issues" with journal impact factors are in Abt [2004, 2006a] and Van Noorden [2010]).  In any case, no further use is made of impact factors in this paper.) My final list had 19 journals.  One of these made its first appearance in 2003 ("*Journal of Cosmology and Astroparticle Physics*") while another, *A&A Supplements*, merged with *A&A* in 2001.  Thus, in this paper, data for *A&A Supplements* will always be merged with *A&A* itself.  Similarly, the numbers for the *Astrophysical Journal* include *ApJ Letters (ApJL).*[3]

---

[1] Some of the journals listed in this category are not primarily astronomy related while other journals that are primarily astronomy related are not included in the JCR list.  Neither of these facts has a significant effect on the analysis in this paper.  Furthermore, neither *Nature* nor *Science* are on JCR's list; they are categorized as "multi-disciplinary".  These two weeklies required "special handling" as will be described later.

[2] The JCR defines impact factor of a journal as follows:  "[It] is calculated by dividing the number of citations in the [JCR year] by the total number of articles published in the two previous years. An Impact Factor of 1.0 means that, *on average*, the articles published one or two year ago have been cited one time. An Impact Factor of 2.5 means that, on average, the articles published one or two year ago have been cited two and a half times."

[3] Until late 2007 *ApJL* was not identified separately from ApJ itself; so that all JCR searches relevant to *ApJ* automatically included *ApJL*.  Web of Science searches also did not distinguish between the main Journal and the *Letters* except for the fact that page numbers for articles appearing in *ApJL* had an "L" before the number.  However, about the time that the AAS changed publishers for the *ApJ* the Web of Science and the JCR began to treat the main journal and the letters as two separate journals.  This, according to Thomson Reuters Customer Technical Support is because the new publisher, IoP, assigned different ISSN numbers to the two.  So when Thompson Reuters receives the data for the journal from IoP it comes in as two separate journals and is entered as such in their data base.  So a Web of Science user must request information for both *as distinct journals*.

Table 1 lists 17 of the 19 journals from which the top 100 articles are selected along with summary statistics from ISI for 2000 – 2008; *Nature* and *Science* are not listed since they have to be treated differently – see below. There are no entries for 2009 because JCR had not yet compiled the statistics for that year when I did the citation search. Also, the ISI's JCR online database is missing the integrated data for *Advances in Space Research* for 2006. For each year the two columns give the number of citations to all articles published in that journal for that year (as of early 2010) and the number of those articles. The sum over all journals in Table 1 for each year is a pretty constant fraction of the *total* number of articles published by *all* journals listed by JCR in the category "astronomy and astrophysics", namely between 84 and 89% except for 2000 when it goes up to 92%. Not surprisingly, the journals in Table 1 include most of those in Abt's (2006) list.

*Nature* and *Science* are classified as "multi-disciplinary" by ISI rather than in the astronomy and astrophysics category so they required special handling. I used the "General Search" on ISI's Web of Science pages and went year by year for each of the two journals, and a list of topics that I determined with some trial and error. [4] As an additional aid in minimizing the number of "spurious" results, I further restricted the search to specific types of "documents", viz. articles, database review, letters (but *not* research letters in the sense that *Nature* uses the term), notes, and reviews. Once this search was done I then inspected each list to toss out non-astronomical articles, book reviews, etc that made it through the filters.

Two of the journals in my final list are devoted to review articles, rather than articles presenting new findings – *Annual Review of Astronomy & Astrophysics*, and *Space Science Reviews*. These have been retained for this study since this is a survey of *all* the most influential articles.

### 3.1.2. Selecting the 100 Most Cited Papers

The next step was to identify the "greatest hits", the 100 most cited articles for each year. Since the *ApJ* always has the lion's share of citations, I used the Cited Reference Search on ISI's Web of Science to first select the 100 most cited papers for each year from the *ApJ*. This served as an absolute lower limit for the selection of the 100 most cited from all journals. Then for each of the other journals in Table 1 and for each year I selected those papers that had at least as many citations as the paper that ranked 100 on the list for the *ApJ*. In spite of the fact that the journals in Table 1 have already been selected from the full list of journals, nearly half of them never had any papers with as many citations as the 100[th] ranked paper on the *ApJ* list.

Table 2 shows the distribution amongst the journals for the top 100 most cited papers in the sample. The last line, "Other", is for journals that generally had an average of 1 or less top-

---

Unfortunately, JCR does not input data for a "new" journal until there is a few years worth of data. Thus any data extracted from JCR for the *ApJ* after late 2007 applies *only* to the main journal. Thus, for 2007, JCR shows 2796 articles in the main journal; while the Web of Science lists 52 *additional* articles in *ApJL*. For 2008 the corresponding numbers are 2128 and 677, respectively. For 2009 they are 2795 and 697, respectively. The present paper continues to treat the two parts of the *ApJ* as one and refers to it simply as "the *ApJ*".

[4] Here is the full string in case a reader wishes to do her or his own search: This string of topics with Boolean operators seemed to be the most effective at both selecting all astronomy related articles while at the same time minimizing the number of non-astronomical articles: *"black hole" or astro\* or cosmo\* or solar\* or planet\* or stellar\* or star or galaxy or galactic or GRB or cosmic or "gamma-ray" or pulsar or Mars or Saturn or Pluto or Titan or Uranus or Neptune or "deep impact" not climate not genetic not neuron\* not human"*.

cited articles per year. This latter group includes Icarus and PASJ. An exception is JCosAPP. Although it had zero or one articles that made the top 100 from 2003 (its first year of publication) through 2008, in 2009 it had 10 articles. This table shows that only 6 journals account for 85% of the 1000 most cited papers over the ten year period. Appendix B gives the $1^{st}$, $50^{th}$, and $100^{th}$ ranked papers for each year.

The numbers in Table 2, though, do not tell the whole story though since the *ApJ, A&A,* and *MNRAS* account for about 75% of *all* articles published in astronomy over the 10 year time period based on the data from ISI's JCR. Table 3 gives the percentages of articles from each journal in Table 2 that make it into the top 100 for that year. For *Nature* and *Science* these numbers refer to just how many articles on astronomy and astrophysics were identified for each year as described previously. As may be seen, the most "successful" publication in terms of fraction of articles published that make it into the top 100 is *ARAA* with *Nature* second and *Science* third. Both Nature and Science, though, limit the number of astronomy articles that appear in their weekly issues; thus they are already preselecting candidates for the top articles. For comparison, three-fourths or more of the articles submitted for publication to the *ApJ* (including *Letters*), *AJ, A&A,* and *MNRAS* eventually appear in those journals.[5] For reference, the average over 10 years of all articles published in these journals that make it into the top 100 is 1.4%. This indicates that on average *A&A* and *MNRAS* are relative under performers while *AJ* and *ApJ* are about average. Low numbers for Solar Physics may in part reflect the relatively small size of the solar community. The same is probably true for *Icarus*. Bear in mind, though, that most of the small percentages have significant standard deviations (last column of Table 3).

With the procedure I followed there is still the possibility that some of the journals not searched, i.e. not in Table 1, have articles with a high enough citation count to place them in the top 100 for the year. To investigate this possibility, I went back to check if a sampling of the non-Table 1 journals had any articles that would fall in the already selected top 100. Following a suggestion by V. Trimble, I searched *Acta Astronomica, Astronomy Reports, Astronomy Letters, Astronomische Nachrichten, Journal of Astrophysics, Revista Mexicana Astronomy & Astrophysics, Observatory,* and *Baltic Astronomy*. Very few of these journals had any articles with citation counts that came even within a factor of two of the last ranked article in my top 100 lists for each year.

However, 4 articles were found over the 10 year period that had more citations than the $100^{th}$ ranked one for that year after the selection was made. There were two in *Acta Astronomica* in 2002 with citation counts of 210 and 192. These numbers would have placed both in the lower half of the top 100 for that year. Both presented results of large-scale surveys, one an all-sky survey for variable stars, the second the results of the 2001 campaign of OGLE, the Optical Gravitational Lensing Experiment. Most of the same issue of *Acta Astronomica* contained results from OGLE. The other two articles that would have made it into the top 100 are both from *Astronomische Nachrichten* and both have to do with technical aspects of the Sloan Digital Sky Survey (SDSS). They were from 2004 and 2006 with 250 and 158 citations, respectively; so would rank about $30^{th}$ in the highly cited list for their years. All four of these articles were found after the analysis presented in this paper was completed. Since they would constitute only 0.4% of the 1000 articles that are the basis of the analysis, I did not redo the analysis to include them.

---

[5] In the case of the *ApJL*, about 60% are accepted into the Letters while another 15% get moved to the main journal.

### *3.1.3.  Further Comments on Table 2 and Table 3*

As pointed out earlier, since *Nature* and *Science* are multi-disciplinary journals, I needed to use the ISI search engines to select, based on keywords, a preliminary list of articles that likely dealt with astronomical topics for each year.  I then went through these lists usually by title but occasionally with the help of the abstract, to excise any remaining non-astronomical articles.  Thus the total count per year for these two journals may be biased in undercounting astronomy related articles.  With these caveats in mind, for 2007 – 2009 these journals each published about 50 astronomy related articles, or one a week on average.  In 2005 and 2006 Nature had 90 and 72 articles, respectively, while *Science* had 67 and 63, respectively.  For *Nature* these two years were also ones with more than 20% of its astronomy papers making it into the top 100 (see Table 3).  For both journals these numbers dropped down into the 30s for 2000 to 2002.  For 2005 and 2006 both journals had a considerable number of papers presenting results from the Cassini mission to Saturn and Titan, the various missions to Mars, Voyager 1, and Deep Impact.  *Nature* had several Voyager 2 papers in the top 100 in 2008.  For 2000 the three *Science* papers that made it into the top 100 were all from the Mars Global Surveyor.

The *ApJS* has a marked upward spike in 2003.  This was previously pointed out by Abt (2006a; see also Trimble and Ceja 2008) who attributed it to "the extremely high citation rates … for 13 papers in the special issue devoted to the *Wilkinson Microwave Anisotropy Probe*".

The small up tick for *PASP* in 2003 is due to a series of articles that give overviews of the Spitzer legacy projects.

As a partial check on long term trends I carried out identical searches for the top 100 articles for 1990 and 1995.  The last two columns of Table 2 show the distribution amongst journals for these years.  Within the scatter there is not much change going from these two years to the decade sample, although several journals may show a decline in their contribution to the top 100 while a couple of others  may have gone up a bit.  Two journals that now make small contributions to the top 100 but did not even exist in the 1990s are JCosParAp and Astropart. Phys.

### 3.2.  The Time Evolution of Citation Rankings Based on a Comparison of Two Samples

Since the six-100 and ten-100 samples were assembled about two and a half years apart, a comparison of overlapping years should indicate something about the time evolution of the rankings of articles by citation numbers.  A related examination of the lifetimes of astronomical papers was made by Abt (1998) and Abt and Boonyarak (2004).

Data for the six-100 sample were assembled in late 2007 while data for the ten-100 sample were assembled in early 2010.  The second column of Table 4 lists the number of papers in the top 100 samples for the years given in column one that were in the earlier sample but did not make it into the later one, having been replaced by other papers.  The third and fourth columns give the average rankings of the papers that were dropped (Six-100 sample) and those that replaced them (Ten-100 Sample).  This table shows that the number of non-overlap papers declines quickly with age.  Also that the papers that dropped out of the six-100 sample are generally ranked in the bottom quarter of the sample, while the new ones that came in, although still in the bottom half of their ten-100 samples, have, on average, systematically higher rankings than those that they replaced. These results are not surprising.  As papers age, their citation ranking will generally stabilize. And it should be much easier to dislodge a paper from near the bottom of the pile than one in the top half.

Are there any trends apparent in the samples of non-overlap papers? First, I broke them down by journal – noting those journals that lost papers from the six-100 sample and those that gained papers in the newer, ten-100 sample. These results are given in Table 5 where a negative entry indicates a loss of articles. The last two columns give the arithmetic sum for each journal and its standard deviation. Based on these numbers there appear to be some statistically significant winners and losers. The *ApJ* and *MNRAS* are the losers while the winners are *A&A*, *ARAA*, and *PASP*.

There is an interesting scientific thread in the lost and gained samples. Examination of the article titles revealed a strong bias as to topic. For 2001 to 2003, while about 11 of the 13 or 14 papers lost *each year* (see Table 4) covered pretty much the entire electromagnetic spectrum, they *all* dealt with extra-galactic topics while only three to five of the new papers that replaced them were extra-galactic. For 2004 amongst the "lost" papers were 11 extra-galactic ones and 5 from the Martian Rovers while the 2004 "new" papers had 8 extra-galactic ones and only one about Mars. For 2005, the "lost" and "new" samples had 20 and 10 extra-galactic papers, respectively. Finally, for 2006, of the 42 lost papers, 33 were extra-galactic to be replaced by 32 other extra-galactic ones.

Overall this suggests that extra-galactic papers have a more rapid rise, about one to three years, in their citation counts than papers on other research areas. After this initial rapid rise, the rate at which extra-galactic papers accumulate citations drops down to or below that of the papers in the other areas. Furthermore, the lost papers appear to be dominated by observational studies whereas the new ones are more heavily weighted towards theory and instrumentation. These findings are consistent with Trimble and Ceja's (2008, 2010) that the rate at which citations accumulate can be a strong function of a number of different variables, particularly the field of research, and Abt's (1996, 1998) observation that the half lives of papers can vary considerably. Trimble and Ceja also note that overall the rate at which citations accumulate, especially in the first few years after a paper's publication, is much steeper than linear. All of these issues will be considered more fully in a subsequent paper.

## 4. TIME DEPENDENCE OF NUMBER OF AUTHORS PER ARTICLE: THE INCREASING PREVALENCE OF TEAMS IN ASTRONOMY

Wuchty et al (2007) on the basis of an analysis of nearly 20 million papers over 5 decades demonstrate that "teams increasingly dominate solo authors in the production of knowledge". They also show that research papers produced by teams are more frequently cited than those by individuals, that this disparity has been increasing with time, and that "exceptionally high-impact research" is now being produced by teams whereas in the past, they claim, such influential research was once the almost exclusive domain of researchers working alone. Wuchty et al.'s investigation subdivided fields into three broad categories, one of which, science and engineering, encompassed 171 different fields. With the information assembled for the present study we can see if such a trend is present for astronomical research papers over a much shorter time frame

This section will examine the time dependence of the total number of authors and the total number of unique authors for the top 100 papers for each year as well as an examination of the distribution of number of papers vs. number of authors for each year.

Table 6 lists, in the second column, the total number of authors for the top 100 papers for each year. The third column of the table gives the number of *unique* authors on these papers for each year[6]. Both of these numbers are illustrated by the bars in Figure 1 (scale on the left). This figure illustrates one of the main findings of this paper, namely that *the total number of authors for the top 100 papers has more than tripled between 2000 and 2009*. The rise is steep from 2000 to 2007 and levels off from 2007 to 2009. Since the ISI data base does not give the number of authors per article, I could not easily determine if such a rise is typical for *all* astronomy related articles or is confined to the top 100 for each year. From the numbers for 1995 and 1990 in the table we see that the rise over the past decade appears to be considerably steeper than that which occurred over the decade of the 1990s for the top 100 papers.

The number of unique authors has also increased strongly over the past decade, though not quite as rapidly as the total number. The number of unique authors tripled from 2000 to 2008 but fell in 2009 both absolutely and as a fraction of the total number. We also note that in 1990 and 1995 about 90% of all authors were unique, i.e. an individual appeared on only one of the top 100 papers in those years. Overall, the evidence from Figure 1 is that over the past two decades the number of unique authors for the top 100 papers in astronomy for each year has declined from about 90% in 1990, to about 60% at the end of the last decade, i.e. now on average an individual author appears on nearly two of the top 100 papers.

Figure 1 also plots membership numbers for the AAS and IAU (right hand scale). These numbers have stayed flat for the AAS while there has been an increase of about 20% for the IAU over this same time period. So we conclude that that large increase in number of authors for the

---

[6] To get the number of unique authors I first eliminated all identical names including initials from the list of all authors for the top 100 papers for each year. Then for authors with the same last name but different initials, I eliminated those entries that appeared to be the same except for a missing middle initial. This procedure will tend to undercount unique authors since some may have no middle initial but the same first initial as another author with a middle initial. Random checking indicates that the uncertainty in the numbers of unique authors is at most a few percent.

top 100 papers is unrelated to the number of astronomers, at least as measured by membership in the AAS or IAU (cf Abt 2007). Interpretations of this result are in Section 7.2

The Xs in Figure 1 shows the total number of articles published per year for all of the journals listed in Table 1 as given on the last line of that table. The number of articles for 2009, 9205, was determined from the listings for the individual journals since, as pointed out earlier, JCR had not yet posted compilations for 2009. The scale for this quantity is also on the right side of the figure. As noted in Section 2 these numbers represent between 84 and 89% of all articles published in *all* journals that JCR considers being in the area of astronomy and astrophysics (these percentages can be derived directly from the entries in the JCR database). Thus while we obviously cannot connect authors to society members with the data in hand, we can state that the ratio of number of top-ranked papers to number of IAU members is on average close to or a bit under one in these journals while for AAS members it is about 20 to 30% greater than one. In this regard it is worth pointing out that many members of the AAS are students so would have a very limited publication record thus increasing the latter ratio further. At the same time, the IAU does not welcome student members. These ratios are quite constant over the 10-year interval. This is discussed further in Section 7.2.

Next we ask how does the rate of increase (or decrease) with time of the number of papers with a given number of authors depend on the latter number. Since we are dealing with a fixed population of 100 papers per year there have to be winners and losers. For each year we have counted how many of the top 100 articles have "X" number of authors, where X goes from 1 up to the maximum number of authors that any article had in that year. To smooth the statistics I have taken two year averages. Figure 2 shows the cumulative number of papers with number of authors equal to or less than the X value. With the exception of the two most recent years there is a strong and systematic trend with time. Older papers with a small number of authors are far more common than newer papers with the same small numbers of authors. For example, for 2006 and 2007 50% of the papers in the top 100 had 12 or more authors. In 2000 and 2001 only 18% had 12 or more authors while the average for 1990 and 1995 indicates that only 9% of the top papers had 12 or more.

Now we want to investigate the shift in frequency of number of papers with X number of authors amongst the top 100. For each number between 1 and 30 authors per paper I calculated the dependence on time for 2000 to 2009 of the number of papers in the top 100 with the given number of authors. It was not feasible to do this for more than 30 authors as the number of papers become too sparse. Figure 3 illustrates the results. This figure shows that papers with 4 or fewer authors have strong negative slopes, i.e. the frequency of such papers declined considerably going from 2000 to 2009. For 5 or more authors on the other hand the trend is small but generally positive. If the years 1990 and 1995 are included in the slope calculation (open circles in Figure 3) the scatter is reduced; now all papers with 5 or more authors shows a positive slope with time.

We can gain further insight as to how the number of authors of papers that are in the top 100 varies from year to year by looking at quartile values. These values in Table 6 are the quartiles for each year's 100 top papers arranged not by number of citations per paper but in increasing order by the number of authors per paper. The sense of the quartiles is such that the number of authors for 25% of the papers is equal to or less than Q1, for 50% of the papers it is equal to or less than Q2, etc. Q2 is, obviously, the median of the number of authors per paper for each year while Q4 is the maximum number of authors on a paper for that year. From 1990 through 2002 the upper bound to the median, Q2, is about 4.0 authors per paper, while from 2006 onwards only 25% or less of the papers have such a small number of authors. This is just another expression of the result seen in Figure 2. The leveling off in the total number of authors for each year's top 100 papers that is apparent in Figure 1 for 2007 through 2009 is also reflected in the average of the cumulative distributions for these two years (Figure 2) and in the decline in the Q1 and Q2 values for them (Table 6).

Table 6 suggests that *relative* changes of the quartile values with time are similar to one another and to the relative change in the total number of authors per year. In order to illustrate this I have normalized the numbers of authors per article for each year by the total number of authors for that year and recomputed the quartile values. This puts all values on the same scale and makes comparisons easier. Figure 4 shows the trends in the normalized quartile values as well as for the total number of authors for each year for the top 100 papers. Keep in mind that we are dealing with the same number of papers for every year.

In Figure 4 first note the similarity of the distributions from 2000 through 2006. Even the lowest quartile of normalized values with the smallest number of authors per paper has increased by a factor of three over the past decade. The un-normalized number of authors that defines Q1 (see Table 6) varies from two to five; for Q2 it goes from 4 to 10 or 15; for Q3 from 6 to 25, while the total number of authors goes from 800 to nearly 3000 over this ten-year time span. The conclusion, then is that from 2000 through 2006 the overall dependence of number of papers published by X number of authors did not change shape significantly but rather it just shifted towards a greater number of authors per paper. Based on just two years of sampling for the decade of the 1990s, it would appear that these time dependences were not as steep then as for the decade that just ended.

Abt (2000d) finds an average of 3.8 authors per paper for the *ApJ* for the last decade of the 20[th] century. Stanek (2009) study, based on all papers published by ApJ, ApJS, AJ, A&A, and MNRAS between 2000 and 2004, inclusive, found a median of 3 authors for the 30,000 papers in his sample. The average number of authors here for the top 100 cited papers in 2000 is 8.0, more than twice as many as Abt's number for the preceding decade. Table 6 indicates that the median number of authors for the top 100 papers (the Q2s in the table) was 7 in 2004, also more than twice Stanek's value for the preceding several years. These new results are consistent with the finding discussed below that highly cited papers have on average a greater number of authors than less cited ones.[7]

Abt (2007b) has also studied the frequency of single author papers in astronomy and three other fields of science. For astronomy he considers four journals – *ApJ, AJ, MNRAS,* and *A&A.* For 2000 and 2005 he finds that the average frequency of occurrence of single author papers in these journals is 11.7 and 10.3% respectively, for the two years. Stanek (2009) based on his 2000 to 2004 sample noted just above, derives a similar frequency of 10% for single author papers. For the present sample, 8% of the top 100 papers in 2000 in Abt's sample of 4 journals are single author ones, while for 2005 3% of the top 100 papers are single author. Overall, for the first half of the last decade, 2000 to 2004, 6% of the top 500 papers are single author ones; for the last half of the decade only 2.8% are.

---

[7] The median number of authors for the top one hundred papers for 2000 is only 4.0. That this is only half of the average is not surprising given the skewness of the distribution of the number of authors per paper.

One might think that since the list of journals searched for the top 100 papers includes two publications devoted to invited review articles (ARA&A and SpSciRev), that these would dominate the single author listings. They certainly do contribute a disproportionately large share of single author articles: There are 44 single author papers in the top 1000 for 2000-2009, 6 of which or 15% are from these two publications; compare this to the 5% they contribute overall to the 1000 top papers. However, even the single author papers in the top 1000 sample are dominated by ones appearing in the large journals. For example, 59% of the 44 single author papers are from the ApJ and MNRAS, whereas these two journals account for 53% of the 1000 papers in the top 100 for the same time period.

Abt (2003) comments on papers with large numbers of authors and finds that "in astronomy the large number rarely exceeds 25". Stanek (2009) in his 30,000 paper sample from 2000 to 2004 finds that only 100 of these, or 0.3%, have 50 or more authors. In contrast, for the 1000 papers studied here, 18% have 25 or more authors, 6.5% have 50 or, and 3.5% have 100 or more authors. To more directly compare with Stanek's sample, between 2000 and 2004 22 of the top 500 papers had 50 or more authors, or 4.4% - a frequency more than an order of magnitude greater than that found by Stanek. As Figure 1 and Figure 2 imply, these numbers have been increasing steeply over the course of the past decade. For example from 2005 to 2009 60 of the top 500 papers had more than 50 authors, or 12%, nearly triple that for the previous 5 years of "greatest hits". If this trend continues, astronomy may need to look to the example set by the high energy physics community in interpreting the meaning of authorship when such large numbers of authors are involved.[8]

---

[8] The increasing prevalence of group research in physics over the latter half of the 20th Century has been commented on by Kevles (1995, esp. pp. 374 and 389).

## 5. THE TIME DEPENDENCE OF THE NUMBER OF CITATIONS
## AS A FUNCTION OF RANK

Not only is the total number of citations to a paper a function of time, but as we showed in a previous section even the citations rankings of papers can change significantly year to year. We will now investigate the functional dependence of the number of citation upon rank since there is no *a priori* reason for this to be the same from year to year.

In this section we will procede as follows: Each paper for each year is assigned a citation rank for that year based on the number of citations to that paper as of early 2010 and as tabulated by ISI. Then the absolute number of citations received by each of the top 100 papers in a year is normalized by dividing that number by the sum of citations received by all 100 papers as of early 2010. This normalization removes the fact that citation counts increase with time and allows the *distribution* of citations from each of the ten years to be compared. Finally, to avoid having to deal with numbers much smaller than one, the normalized citation numbers are multiplied by 1000.

Figure 5 shows that the *shapes* of the distributions defined by the normalized number of citations versus paper rank for each year are nearly identical for the ten years being considered with very little scatter except for the top 30 or so and that for rankings higher than about 10 all of the counts rise steeply. This behavior is shown in more detail in Figure 6. The red line in Figure 6 is the average of the normalized number of citations for each rank over ten years. This average line is repeated in Figure 7 for all 100 ranks while the vertical bars in this figure are the standard deviations of the ten data points at each rank. These are typically 3% or less for ranks 40 and greater, rising to 7% at rank 20 and to 15% at rank 7. From rank 6 to 1 the dispersion increases quickly. This is better illustrated in Figure 8, which expands the distribution for the topmost ranks. Again the red line is the average value.

For ranks 6 to 100 the average values are fit nearly exactly by a power law with exponent equal to –0.436 and an $R^2$ value of 0.994.[9]  Over this range the power law is essentially identical to the average line as may be seen in Figure 6 and Figure 8 where the black lines are the power law fit and its extrapolation to rank 1. For most of the data in the 2000 to 2009 period the extrapolated power law closely approximates the data. The significant scatter for some of these most highly ranked papers, though, is always in the sense of too many citations for the ranking. The greatest deviations are for the top 5 points for 2009 and the top 3 for 2003, all of which lie well above the power law fit. Some of the top 3 points for 2006 and 2000 also appear to be high, and the first ranked point for 2007 is high. What can be said about these papers that have an inordinately large number of citations?

Papers published in 2009:  Four of the top 5 papers published in 2009 present the successive data releases of observations by the WMAP all of which appeared in *ApJS*. The fifth paper in the top five (ranked 4th) appeared in *Nature* and is concerned with observations of an anomalously large number of cosmic ray positrons that may have resulted from the annihilation of dark matter. The somewhat high 10 to 12 points in the 2009 data (Figure 6) refer to two more

WMAP and cosmic ray positron papers as well as the discovery paper of an extra-solar planetary system with three super-Earths.

Papers published in 2007: The first ranked 2007 paper which has nearly six times as many citations as the second ranked one for that year is the three year analysis of WMAP data and its implications for cosmology. Two more of the top five papers for 2007 also discuss three-year WMAP data but as Figure 8 shows, all of the other most highly ranked papers for 2007 fall on or close to the extrapolation of the power law fit to the 6 to 100 data points.

Papers published in 2003: Two of the top three papers for 2003 are the first year results from WMAP. The third paper presents the details of what is considered to be the standard model for computing the spectral evolution of a stellar population by Bruzual and Charlot (2003). This is the only paper amongst those with an anomalously high citation rate that is not based on observational data. The number one paper on WMAP has 7 times as many citations as the number 4 ranked paper for 2003, about the cosmological results from the high-z supernova search.

The 2006 and 2000 data: The top three publications for 2006 appear to be a bit high. The number 1 paper is an overview of the Two Micron All Sky Survey (2MASS), while the number 2 paper presents first year data from the Supernova Legacy Project on Omega and $w$; the number three paper is about AGNs. The top and second ranked papers for 2000 which may also be a bit high are, respectively, a technical summary of the Sloan Digital Sky Survey and an interpretation of the CMB observations.

So of the nine most deviant points amongst the top 5 papers in years 2000 to 2009, seven present results from WMAP. One of the remaining two is a "one off" observation of an event that may be linked to dark matter, while the other is a modeling paper with very wide applicability rather than presenting a new result or interpretation. All of these papers with the exception of the one on stellar models are also in the top quartile of number of authors for the years in which they were published. Finally, nearly all of the papers noted as being definitely or possibly high are based on large observational surveys from space and the ground.

# 6. OTHER CORRELATIONS

## 6.1. Number of Pages per article versus Time and Citation Rank

Abt (2000b, 2000d, 2003) has examined the growth in the length of articles in the three major American astronomical journals – *ApJ*, *AJ*, and *PASP* and that towards the end of the 20[th] century the average length of a paper was asymptotically approaching 12-14 1000-word pages. A random sampling of 10 issues of these three journals published between 2000 and 2009 yields an average of 11.9 pages per article with no obvious time dependence. This is probably consistent with Abt's 12-14 given that 11.9 is a straight average, not normalized to 1000 word-pages. On the other hand, for the sample of 1000 highly cited articles in this study, I find an average length of 18.7 pages with a non-statistically significant increase in page length of 8% over 10 years. For just articles from *ApJ*, *AJ*, and *PASP* which account for 50% of the 1000 top articles, the average page length is 18.4, not significantly different from 18.7. Thus another important finding of this study is that the most highly cited articles in refereed astronomical journals are typically one and a half times longer than the average of all articles and are not dependant on time over a ten year period.

For the control years 1995 and 1990 the average number of pages for the top 100 articles in *ApJ*, *AJ*, and *PASP* is 17.0, marginally lower than the 18.4 average noted in the previous paragraph, but still well above the average for *all* articles in these three journals found by Abt.[10]

Is there a dependence of page length on citation frequency amongst the highly cited sample studied in this paper? Since there is no statistically significant dependence on year of the *total number* of pages for the top 100 articles from 2000 to 2009, we average by citation rank over the ten years. Figure 9 illustrates the dependence of the average number of pages per article versus the citation rank of the paper, where "average" is that of the ten papers at each citation rank for the ten years. The slope of the linear regression (the solid red line) is −0.029 with an $R^2$ of 0.044. The 95% confidence limits are indicated by the dashed lines. Thus, on average the top ranked papers in the present ten-year sample have 2.9 more pages per article than the bottom ranked papers from the same sample.[11] This result supports Abt's (1998) contention that on average longer papers are cited more frequently.[12] This is discussed in Section 7.2.

---

[10] Interestingly, the average number of pages per article for *all* 100 top articles in 1990 and 1995 is 21.1 and 25.8, respectively, significantly higher than the averages for 2000 to 2009. However, if I exclude *ARA&A*, *SpSciRev*, and *ApJS* then the averages are 15.7 and 19.0, respectively, drops of 5.4 and 6.5 pages. In contrast, for a sampling of years between 2000 and 2009, if I exclude these three publications from the page totals the drop is typically only 2 pages per year. Thus it appears that for the years 1995 and 1990 the *ApJS*, *ARA&A*, and *SSRv* had significantly longer articles than was typical for the decade just ended.

[11] A small but non-significant negative slope was also found for the two year 1990 and 1995 samples. Given that there is only one-fifth as many data as for the 2000 to 2009 sample, the scatter is much greater.

[12] Stanek (2008), based on his sample of 30,000 papers referred to early in the current paper, finds that papers of more than 80 pages are cited less frequently than somewhat shorter papers. The present data do not support Stanek's finding; this is probably because the present sample of papers is so different from his. Furthermore the data here do not give any evidence of a local maximum in the number of citations for papers four pages long, again probably because of the differences in the nature of the samples.

## 6.2.    Number of Authors per Article versus Citation Rank

Next we investigate the correlation of citation rank and number of authors.  Figure 1 clearly shows the strong time dependence of the *total* number of authors per year for the top 100 papers.  To remove this time dependence I have normalized the number of authors of each of the 100 articles for each year as described in Section 4 with  the same procedure as that used for citation numbers.  The result is shown in Figure 10.  The steep upturn in average number of authors for average citation rank higher than about 20 is similar to that for citation counts versus rank.  In fact, the data in Figure 10 are well fit by a power law (in red) with exponent -0.408 whereas the citation count vs rank power law fit had an exponent equal to -0.436.

The dependence of average number of authors on citation ranking can also be examined by determining the *ranking* by author count and plotting that against citation ranking.  This former quantity is the average number of (normalized) authors for papers of a given citation ranking just derived and put into rank order by number of authors with a rank of 1 for the largest number of authors.  The result is illustrated in Figure 11.  Note the concentration of points in the lower left corner.  Seventy percent of the points with an average 10-year citation ranking in the top 20 also have a number of authors ranking in the top 20; the converse is true as well.   Thus we conclude that on average for the 2000 to 2009  period the highest ranked papers in terms of citation counts are also the highest ranked in terms of number of authors; the inverse is also true.[13]

## 6.3.    Total Number of Articles

With the data base assembled for this study trends other than those for the 100 most cited articles can be investigated.  Here we examine one such trend – the total number of articles published per year.  A cautionary note: changes in number of articles published per year can be due to a combination of causes unrelated to science.  For example a slow down in the rate at which articles go through the whole publication process, a conscious effort on the part of the journals to restrict the number of articles that will be published, financial issues, etc.

Abt (2000b, 2000d) found that in the three American journals, *ApJ*, *AJ*, and *PASP*, starting in the 1960's, the number of papers published per year began to increase linearly by 71 papers per year through the end of the $20^{th}$ century. From his (2000d) Table 1 this corresponds to 2.7% per year for the decade of the 1990s on average.  Has this trend continued into the first decade of the $21^{st}$ century?   From the values for these three journals from 2000 to 2008 (Table 1) we derive a slope of 49 articles per year ($R^2 = 0.63$) over this time span or 1.5% per year.

If we sum over *all* of the journals in the JCR database that are in the category of astronomy or astrophysics, 34 in 2000 and 43 in 2008 (correcting for *ApJL* as described earlier), we find a linear increase of 242 articles per year, or 2.3%, with an $R^2$ of 0.66.  These are the open circles in Figure 12. If we consider only the 13 journals that contribute to the 100 most cited per year, the absolute value of the increase is quite similar - 234 articles per year, but the percentage is of course greater at 2.8% from 2000 to 2008, with a $R^2$ of 0.92. These are the filled circles in

---

[13] Stanek (2009) for his sample of all 30,000 papers published in 5 journals between 2000 and 2004 also found a strong correlation between the number of citations an article receives and the number of authors of the article, although there was a wide scatter.  Basically he found that the median number of citations increased by about a factor of 4 between articles with a single author and those with 50 or more authors.  This is in good qualitative agreement with the present results.  A quantitative comparison is not possible.

Figure 12. The solid lines in the figure show the regression lines. The 13 journal sample includes astronomy related articles from *Science* and *Nature;* as noted earlier these two publications are not in the JCR astronomy or astrophysics database. If articles in *Science* and *Nature* are excluded from the 13 journal subset the linear increase is still 227 articles per year. This subset of 13, only one-third of all of the journals included in the astronomy and astrophysics category by the JCR, still accounts for more than three-quarters of the total number of articles by all of the journals in the category as may be seen in the figure. Taken together, these numbers imply that the 13 (or 11) journal subset accounts for a disproportionately large fraction of the yearly increase in articles.

If the articles published by these same 13 journals in 1995 and 1990, 6092 and 4388, respectively, are included, the calculated yearly increase is lower, 202 per year, suggesting that publication rates have increased over the most recent decade (but see the cautionary note at the beginning of this subsection).

## 7. SUMMARY AND DISCUSSION

This is the first paper in a series that will examine changes and trends in astronomical publications, particularly over the first decade of the 21st century, 2000 to 2009. It is based on astronomy's "greatest hits" for these ten years, the 100 most cited papers for each year. These papers are primarily refereed research ones but about 5% are reviews articles. The top 100 papers from 1995 and 1990 have supplemented this database. The basic characteristics of these 1200 papers *qua* papers have been extracted from the ISI web site and analyzed.

### 7.1. This Paper's Main Findings

The main findings that have resulted from this analysis are:

*Journals in which the papers were published (Table 1, Table 2, and Table 3):* Between 34 and 43 journals are classified as "astronomy and astrophysics" on the JCR website (the exact number has increased with time and does not include *Nature* or *Science*). The same 5 journals (*ApJ, ApJS, AJ, A&A,* and *MNRAS*) account for between 80 and 85% of the *total* number of citations for each year between 2000 and 2008. Averaged over the 10 years from 2000 to 2009 these same journals account for 77% of the 1000 top papers. But these five also account for between 70 and 80% of all astronomy articles published during any given year. Thus as a group they are not outstandingly successful, just outstandingly prolific.

*The staying power of papers in the top 100 (Table 4 and Table 5):* The years 2001 to 2006 were sampled for the top 100 articles twice, about two years apart. Between 13 and 42% of the papers in the earlier sampling did not appear in the later sampling. The number of dropouts varied strongly and systematically with age – the most recent year (2006) had the most dropouts, 42 out of 100, while the oldest two years (2001 and 2002) had the fewest, 13 of 100. The dropouts were strongly biased towards papers relating to extra-galactic research and were replaced by non-extra-galactic ones. This bias is more evident for the years older than 2006. This result suggests that papers on extra-galactic topics experience an initial rise in accumulating citations that is more rapid (1 to 3 years) than that for papers on other topics (3 or more years) but afterwards the rate of accumulation drops to be comparable to or less than that for the papers on other topics.

*The total number of authors for the top 100 (Figure 1):* The total number of authors on the top 100 papers for each year from 2000 to 2009 has more than tripled. In the same time period the number of *unique* authors (i.e. counting each author's name only once even if it appears on more than one of the 100 papers) for each year has more than doubled. Going back to the two years sampled from the 1990s, about 90% of the authors'' names were unique compared with about 60% at the end of the last decade. Over the same time period membership in the AAS and IAU has increased only slowly as has the total number of articles published each year in the journals considered in this paper. Unfortunately it has not been possible to cross-correlate authors' names with names of members.

*The frequency of occurrence of multi-author papers (Figure 2 and Figure 3):* Not only has the total number of authors of the top 100 articles increased strongly, but there has been a strong systematic shift in the distribution of these authors such that the relative number of papers with many authors has increased markedly relative to the number of papers with only a few authors. Even in absolute numbers, the number of papers with 5 or fewer authors has shown a strong decrease with time whereas those with 6 or more authors have generally stayed constant with time or even increased somewhat.

*The time dependence of the number of citations vs. paper rank (Figure 5, Figure 6, and Figure 7):* For the top 100 papers from the years 2000 to 2009 the dependence of number of citations (normalized to remove the overall increase with time) on citation rank is remarkable uniform and with little scatter except for a handful of papers at the very top of the rankings. For ranks 6 to 100 all of the data are fit quite precisely by a power law with exponent = −0.436 and an $R^2$ value of 0.994. The deviant papers all scatter *above* an extrapolation of the fit.

*Papers with exceptionally high numbers of citations (Figure 8):* Of the nine papers that deviate the most from an extrapolation of the power-law fit in the sense of having very high numbers of citations, seven present results from successive releases of data from WMAP, one is a "one-off" paper on cosmic ray observations, and the ninth is a theory paper on stellar models. All of these papers except for the latter are also in the top quartile for numbers of authors in their publication years. These eight observational papers as well as a comparable size group that are also high with respect to the power law extrapolation but to a lesser extent, are all based on large observational surveys from space and the ground.

*The correlation between number of authors and number of citations (Figure 10 and Figure 11):* When the number of authors per article is normalized by removing the strong yearly increase in total number for the year, we find that the average number of authors per article, averaged over 10 years, increases by about a factor of two in going from a citation rank of 100 to a rank of 20. Higher than this the number of authors per article increases sharply. A power law with an exponent close to that found for the dependence of citation counts on citation rank is also a good fit to the overall distribution of number of authors versus citation rank. The correlation between ranking by citations and number of authors is tightest for ranks one through twenty for both: 70% of those with such a ranking for either property also have this high a ranking for the other property.

*The number of pages per article (Figure 9):* The average page length of the 1000 articles in this study is 18.7; it does not appear to be time dependent.[14] About half of these articles are from the three main American journals, *ApJ*, *AJ*, and *PASP*, and have the same average page length – 18.4 pages. This length is about one and a half times longer than is typical for all refereed astronomical articles. Furthermore, even amongst the present sample of the top 100 papers per year, articles at the "top of the top" are on average 2.9 pages longer than articles at the "bottom of the top".

---

[14] This average page length does not take into account extensive tabular material that now is placed on the web by many authors. In the past, articles in many journals, especially the *ApJS*, would have often included such material.

*The time dependence of number of articles published (Figure 12):* The number of articles published in all astronomy and astrophysics journals in the JCR database shows an average increase of 242 articles per year from 2000 (34 journals) through 2008 (43 journals), an increase of 2.3% per year. For the subset of 13 journals that contribute to the top 100 articles per year, the annual increase is essentially the same, 234 articles over the same time period (or 227 articles if Science and Nature are excluded as they are in the JCR database). These 13 journals account for more than three-quarters of all articles published by journals in the JCR total and thus a disproportionately large fraction of the yearly increase in number of articles published.

Finally, a brief comparison of citation counts from ISI and from ADS is presented in the Appendix to this paper.

## 7.2. Discussion and Questions for Further Study

A detailed discussion and interpretation of these findings will appear in a subsequent paper; here I will just make some brief remarks and suggest areas for further investigation.

The past decade has seen a phenomenal increase in usage of the internet. In the introduction to this paper I cited this fact as one reason one why the past decade has been transformative for astronomy. Evidence for the degree to which internet usage has pervaded astronomical research may be seen from Figure 13. This figure shows usage statistics from NASA's ADS kindly made available to me by A. Accomazzi. The number of users has increased by nearly a factor of 10 over the past decade, while the number of abstracts examined has increased by about a factor of 30. And although the full text of papers can be read online, the number of their downloads has increased substantially as well, though not to the same degree as the other two indicators. There are other quantitative measures of the significant increase in web access to astronomical resources. The number of internet searches that have used the tools available at MAST, the Multi-mission Archive at STScI, has gone from about one-million in 2001 to over 19 million in 2009.

I do not think that the sharp rises seen both in number of authors of top ranked papers (Figure 1) and in internet usage (Figure 13) over the same time period is coincidental. The ease and rapidity of communication has greatly facilitated the ability of many people to work together on the same project, including the end product of a research project – the assembly and editing of the resulting papers. And there are good reasons why larger teams of people are required. These include; the international character of nearly all of the major space missions; the increased complexity of modern instruments and the reduction of data that result from them often requiring a variety of "experts"; and the use of telescopes and instruments capable of observing across the electromagnetic spectrum to address a particular question, again requiring researchers with specific areas of expertise. Finally, as discussed below, there are the many large surveys that have been carried out over the past decade that have required the participation of many individuals. But these things are symbiotic – there *can* be more cooperation between groups separated by large distances and international boundaries *because* of the ease and rapidity of communicating via the internet. An interesting question to pursue would be to chart the frequency of international collaborations and its dependence, if any, on citation ranking of papers.

In the course of assembling the data for this paper I noticed that publications with large numbers of authors tend to fall into two distinct and for the most part non-intersecting groups - research projects that involve a great deal of work over an extended period of time, and those that are carried out over a relatively short time span but require coordinated observations with a large array of facilities and equipment. An example of the latter would be follow-up observations to a gamma ray burst. Examples of the former can include massive surveys that require only a limited suite of instruments (e.g. the SDSS) to survey-like programs that require a large suite of facilities, e.g. the study of type Ia supernovae. Of course there are hybrid programs that encompass large scale surveys with many different facilities, an example being the GOODS program.

As will be shown in subsequent papers, more and more large surveys with major impact are being carried out and their results are appearing in the top 100 lists for the years studied. Recall that in Figure 8 most of the points with atypically large citation counts present or are based closely on data from large surveys such as WMAP, SDSS, and 2MASS. These papers are also in the very high author count group. Surveys on this scale usually require large teams of people for a mixture of reasons including overall management of a large and complex project, the complexity of space missions, the significant volume of data that requires handling and interpreting, and the planning for follow-up observations with a broad array of instruments. With the increasing prevalence of large teams it is probably worthwhile to re-examine the issue of self-citations.

Not only does the increasing complexity of modern instruments for telescopes require larger and larger teams to deal with the data, but, especially for those going into space, they also require increasingly large teams of scientists and engineers to design and build them. More and more often all or most of the key people involved in these tasks are included in the author list for papers that result. If membership in astronomical societies is any indictor of size of the field, then consider Figure 1 again. Membership rolls in the AAS and IAU have stayed sensibly constant while the total number of authors has more than tripled. This suggests that either many of the authors of top 100 papers are not members of these societies (which you might expect to be true if many of them are engineers and technicians rather than astronomers) or that many of them are entering the "exalted" ranks of being a co-author on a "greatest hits" paper for the first time. An interesting task to pursue in this regard would be to cross correlate the list of authors with a list of members of one or both societies. It would also be of interest to quantitatively establish how the ranks of scientists doing astronomical research are being added to by an influx of physicists,

Figure 1 raises another interesting question. Although there has been more than a factor of three increase in the total number of *authors* for the top 100 papers, there has been essentially no increase in the total number of *papers* as shown by the Xs in that figure. Furthermore, as is also evident from the figure, it is not just the total number of authors that has increased, but the total number of distinct individuals who are co-authors of these 100 papers has increased as well (the cross-hatched parts of the bars). So a question that could be answered by repeating the analysis done here for a representative sample of all astronomy papers is whether or not a strong temporal increase in the number of authors is present for papers other than the top 100 and, if so, how does this increase depend on citation rank? This paper has already shown that except for the top 20 or so cited papers, there is a significant, though small, positive correlation – the more highly cited papers have more authors. The next paper in this series will address this question.

Why do the top 100 articles have significantly longer page lengths than all articles in the same journals?  Abt (1998) suggested that longer papers produce more citations than shorter ones but did not offer an explanation. While the data in this paper are consistent with longer papers being more highly cited on average (see Figure 9), the implication *is not* that the path to writing a highly cited paper lies in its degree of verboseness.  Rather, a more likely interpretation is that the presentation of important results, especially from large surveys and complex experiments that span big chunks of the electromagnetic spectrum,  and subsequently become highly cited, require an above average number of pages filled with text, tables and figures.  It should also be noted that the page lengths of papers determined from bibliographic data bases do not include the increasingly large amount of associated supplemental material for some papers that is only on the web and not included with the publsihed version.

I am grateful to William S. Smith, President of AURA, for providing support and encouragement for this on-going study of astronomical publications.  He has also given many useful suggestions and pieces of advice. It is a particular pleasure to thank Virginia Trimble for conversations and many suggestions that have substantially improved this paper. Comments and suggestions by Susana Deustua were helpful and appreciated. Alberto Accomazzi, Glenn Miller, and Karen Levay graciously supplied me with data on internet usage. Faye Peterson of the AAS assembled for me the yearly membership totals for that organization and Bob Williams, the IAU President, did the same for his organization.  Helmut Abt commented on a draft of this paper and sent me many reprints of his earlier work that would have been only obtainable on the web for a price.  Anne Kinney supplied me with preprints of her work.  Finally, I thank Jay Gallagher, editor of the *AJ*, for a particularly enlightening conversation.

**APPENDIX A: A COMPARISON OF CITATION COUNTS AS GIVEN BY ADS AND ISI**

The papers that went into the data base for the six-100 and ten-100 data were selected from the *Science Citation Index (SCI)* while as will be explained in a later paper, publications based on big telescope observations were drawn from SAO/NASA's *Astrophysics Data System (ADS)*. Abt (2006b) has noted inconsistencies when comparing citation counts from these two sources. There are 110 papers that belong to both the six-100 and the big telescope samples plus an additional 20 that fell just outside of the top 100 in these samples. For these 130 papers I tested for such systematic differences in citation counts. Table A1, assembled in late 2007, reveals their existence but also suggests that correcting for them is straightforward.

As Abt already pointed out, this table shows ISI systematically undercounts (or ADS over counts) citations with the discrepancy steadily increasing with decreasing age. Although the number of papers for each year is small, the total number of citations to these papers is high and the paper-to-paper scatter is relatively small. Abt's (2006b) Table 1 yields an ADS/ISI ration of 1.15 for 20 papers published in 1997 with his analysis done in 2005, consistent with the time dependence of the ratios in the table here.

I examined a subset of these papers to see if anything stood out that could account for the differences in citation rates. The ADS compilation appears to do a better job than ISI of catching citations that have errors such as wrong page numbers, authors' names mis-spelled, etc. Often in the ISI data base such discrepancies result in two or more listings for the same article. It would require careful checking of the entries for each article in the ISI data base rather than the approach I used, to pick up these variants. Also, as noted by Trimble and Ceja (2010), the ADS listings include many citations from arXiv, astro-ph, and conference proceedings that are not included in the ISI listings. A number of these appear to duplicate articles that later appear in refereed publications, although the ADS website notes that they make every attempt to not have such duplication. Such citations appear to decline with age.

Although Abt also noted that issues with differences in abbreviations used for names of journals are no longer a source of error in the ISI compilations, I found in my 2007 search that for papers published in 2001 it was necessary to search ISI for both "*A&A*" and "*A and A*" for citations to articles in *Astronomy & Astrophysics*. This and related issues are discussed by Sandqvist (2004 and references therein).

I draw two conclusions from the numbers in Table A1: For the few instances in a subsequent paper when I need to compare citation statistics (as opposed to simple rankings) from the two sources, I will correct the ISI numbers to the ADS ones based on the ratios in the table without fear of biasing the results. Second, since the main approach in this paper is to examine *rankings* rather than absolute number of citations, and since I have found that the difference between ADS and ISI counts appears to be independent of the total number of citations within a given year, I will assume that the rankings I use will be unaffected since I *never* mix the two citation counts to get relative rankings for a year.

## APPENDIX B: EXAMPLES OF "GREATEST HITS" PAPERS

Table A2 lists, for each year, the 1st, 50th, and 100th ranked papers together with the statics relevant for the analysis presented in this paper. The first column gives the year of publication and the ranking of the paper. Columns two through 4 give the number of citations a paper has received as of early 2010, its length, and the number of authors, respectively. Then comes a full reference for each paper and finally a few words that characterize the paper's content.

Although this list encompasses only three percent of the 1000 papers in this study, several trends can already be identified which will be the subject of subsequent papers. Of the 30 papers in Table A2, 13 are based on survey data. There are only four papers on astronomical theory, the remainder being on observations work or reviews of such. The 26 observational papers are divided equally between data gathered *primarily* from space *or* the ground; some of these 26 may have data from both sources. Seventeen of the 30 papers are on extra-galactic topics, both observational and theoretical. Finally, four deal with planetary topics including both our Solar System and extra-solar planets.

TABLE 1

FINAL SELECTION OF JOURNALS TO BE SEARCHED:
TOTAL NUMBER OF ARTICLES PUBLISHED DURING EACH YEAR AND THE NUMBER OF CITATIONS TO THOSE ARTICLES AS OF EARLY 2010

| | 2000 | | 2001 | | 2002 | | 2003 | | 2004 | | 2005 | | 2006 | | 2007 | | 2008 | |
|---|---|---|---|---|---|---|---|---|---|---|---|---|---|---|---|---|---|---|
| | Cites | Articles | Cites | Articles | Cites | Articles | Cites | Articles | Cites | Articles | Cites | Articles | Cites | Articles | Cites | Articles | Cites | Articles |
| AdvSpaRes | 2447 | 585 | 2658 | 526 | 2952 | 641 | 3341 | 663 | 3803 | 777 | 3830 | 283 | 162136 | 2707 | 5470 | 430 | 5979 | 549 |
| ARA&A | 4072 | 17 | 3909 | 14 | 5029 | 16 | 4926 | 15 | 5043 | 15 | 5366 | 20 | 5621 | 11 | 5894 | 12 | 6280 | 13 |
| A&A | 33773 | 1343 | 41039 | 1811 | 60646 | 1821 | 63021 | 1936 | 63293 | 1870 | 68577 | 1879 | 70537 | 1935 | 76647 | 1977 | 79218 | 1789 |
| A&A Supp | 6824 | 266 | | | | | | | | | | | | | | | | |
| AJ | 16695 | 513 | 17049 | 533 | 24402 | 500 | 25965 | 485 | 26385 | 523 | 27515 | 449 | 29643 | 480 | 29744 | 448 | 30711 | 415 |
| AstroPP | 1378 | 59 | 1585 | 52 | 1702 | 85 | 1870 | 110 | 2196 | 103 | 2410 | 90 | 2309 | 88 | 2377 | 108 | 2684 | 90 |
| ApJ | 80530 | 2388 | 125102 | 2516 | 141813 | 2299 | 143165 | 2435 | 144264 | 2478 | 152745 | 2700 | 162136 | 2707 | 174555 | 2848 | 184670 | 2805 |
| ApJ Supp | 7387 | 179 | 10756 | 104 | 11338 | 123 | 11869 | 115 | 13565 | 203 | 14962 | 121 | 15781 | 133 | 16656 | 165 | 18111 | 149 |
| AstPhysSpSci | 1838 | 184 | 1725 | 553 | 2316 | 286 | 2394 | 483 | 2584 | 293 | 2416 | 319 | 2868 | 246 | 3256 | 386 | 3869 | 250 |
| Icarus | 6816 | 266 | 7132 | 150 | 8735 | 241 | 8845 | 206 | 8839 | 233 | 10428 | 285 | 10450 | 271 | 11488 | 323 | 12079 | 322 |
| JCosAstPP | | | | | | | | | 1014 | 141 | 1890 | 156 | 2840 | 216 | 4062 | 280 | 5985 | 346 |
| MNRAS | 32750 | 897 | 34598 | 1010 | 40158 | 1073 | 40755 | 1116 | 43858 | 1222 | 47143 | 1194 | 51844 | 1516 | 56193 | 1490 | 61524 | 1567 |
| PlanSpaSci | 3086 | 128 | 3198 | 140 | 3402 | 129 | 3483 | 89 | 3642 | 134 | 3574 | 131 | 3561 | 141 | 4387 | 203 | 4701 | 202 |
| PASJ | 2790 | 133 | 2679 | 133 | 2759 | 117 | 2646 | 118 | 2901 | 111 | 2858 | 97 | 2679 | 105 | 3420 | 179 | 3838 | 123 |
| PASP | 5378 | 154 | 5248 | 128 | 5671 | 114 | 6088 | 115 | 5926 | 111 | 5837 | 117 | 6784 | 153 | 7017 | 118 | 7342 | 116 |
| Solar Phys | 6979 | 199 | 5616 | 163 | 6346 | 126 | 8675 | 168 | 5167 | 102 | 6397 | 134 | 6975 | 177 | 7721 | 168 | 7510 | 200 |
| SpaceSciRev | 2767 | 107 | 2594 | 168 | 3117 | 36 | 3451 | 144 | 3375 | 31 | 3590 | 77 | 3721 | 51 | 4538 | 124 | 4999 | 158 |
| SUM | 215510 | 7418 | 264888 | 8001 | 320386 | 7607 | 330494 | 8198 | 335855 | 8347 | 359538 | 8052 | 377749 | 8230 | 411425 | 9207 | 432401 | 8417 |

TABLE 2

Journal of Origin of the 100 Most Cited Papers by Year
(as of early 2010) from ten-100 Sample

| Journal | 2000 | 2001 | 2002 | 2003 | 2004 | 2005 | 2006 | 2007 | 2008 | 2009 | Total | 1990 | 1995 |
|---|---|---|---|---|---|---|---|---|---|---|---|---|---|
| A&A | 7 | 13 | 7 | 9 | 10 | 11 | 8 | 8 | 8 | 7 | 88 | 9 | 8 |
| AJ | 11 | 13 | 9 | 12 | 8 | 6 | 8 | 3 | 2 | 2 | 74 | 10 | 7 |
| ApJ | 42 | 51 | 42 | 31 | 36 | 30 | 35 | 45 | 42 | 49 | 403 | 40 | 32 |
| ApJ Supp | 3 | 1 | 3 | 13 | 10 | 3 | 7 | 13 | 6 | 15 | 74 | 12 | 12 |
| ARA&A | 6 | 3 | 8 | 2 | 5 | 5 | 4 | 3 | 2 | 0 | 38 | 5 | 4 |
| MNRAS | 13 | 14 | 16 | 18 | 15 | 9 | 16 | 9 | 11 | 8 | 129 | 10 | 7 |
| Nature | 3 | 3 | 6 | 6 | 7 | 19 | 16 | 6 | 10 | 4 | 80 | 6 | 7 |
| PASP | 4 | 1 | 2 | 6 | 2 | 3 | 0 | 1 | 1 | 1 | 21 | 3 | 6 |
| Science | 7 | 0 | 5 | 1 | 3 | 8 | 5 | 1 | 8 | 3 | 41 | 1 | 1 |
| Solar Phys | 0 | 0 | 1 | 0 | 0 | 0 | 0 | 3 | 4 | 1 | 9 | 7 | 7 |
| SpaceSciRev | 0 | 1 | 0 | 0 | 1 | 4 | 0 | 1 | 4 | 0 | 11 | 7 | 7 |
| Other | 4 | 0 | 1 | 2 | 3 | 2 | 1 | 7 | 2 | 10 | 32 | 2 | 2 |

TABLE 3

Percentage of *All* Articles Published by a Journal That Make it into the Top 100 Sample

| Journal | 2000 | 2001 | 2002 | 2003 | 2004 | 2005 | 2006 | 2007 | 2008 | 2009 | Mean | Stdev |
|---|---|---|---|---|---|---|---|---|---|---|---|---|
| A&A | 0.5% | 0.7% | 0.4% | 0.5% | 0.5% | 0.6% | 0.4% | 0.4% | 0.4% | | 0.5% | 0.1% |
| AJ | 2.1 | 2.4 | 1.8 | 2.5 | 1.5 | 1.3 | 1.7 | 0.7 | 0.5 | | 1.7 | 0.7 |
| ApJ | 1.8 | 2.0 | 1.8 | 1.3 | 1.5 | 1.1 | 1.3 | 1.6 | 1.5 | | 1.5 | 0.3 |
| ApJ Supp | 1.7 | 1.0 | 2.4 | 11.3 | 4.9 | 2.5 | 5.3 | 7.9 | 4.0 | | 4.3 | 3.3 |
| ARA&A | 35.3 | 21.4 | 50.0 | 13.3 | 33.3 | 25.0 | 36.4 | 25.0 | 15.4 | | 28.6 | 12.0 |
| MNRAS | 1.4 | 1.4 | 1.5 | 1.6 | 1.2 | 0.8 | 1.1 | 0.6 | 0.7 | | 1.1 | 0.4 |
| Nature | 7.9 | 9.1 | 16.2 | 11.5 | 14.3 | 21.1 | 22.2 | 11.8 | 19.6 | 7.8 | 16.1 | 5.2 |
| PASJ | 0.0 | 0.0 | 0.9 | 0.0 | 0.0 | 1.0 | 0.0 | 3.4 | 0.0 | | 0.7 | 1.1 |
| PASP | 2.6 | 0.8 | 1.8 | 5.2 | 1.8 | 2.6 | 0.0 | 0.8 | 0.9 | | 1.8 | 1.5 |
| Science | 18.4 | 0.0 | 16.1 | 3.0 | 6.5 | 11.9 | 7.9 | 1.9 | 16.3 | 6.0 | 9.1 | 6.9 |
| Solar Phys | 0.0 | 0.0 | 0.8 | 0.0 | 0.0 | 0.0 | 0.0 | 1.8 | 2.0 | | 0.6 | 0.8 |
| SpaceSciRev | 0.0 | 0.6 | 0.0 | 0.0 | 3.2 | 5.2 | 0.0 | 0.8 | 2.5 | | 1.2 | 1.9 |

TABLE 4

THE PAPERS THAT ARE IN ONLY ONE OF
THE SAMPLES

| Year | Non-overlap Papers | Average ranking of non-overlap papers | |
|------|------|------|------|
| | | Ten-100 | Six-100 |
| 2001 | 13 | 79 | 85 |
| 2002 | 13 | 66 | 78 |
| 2003 | 14 | 77 | 82 |
| 2004 | 18 | 68 | 77 |
| 2005 | 24 | 72 | 74 |
| 2006 | 42 | 65 | 71 |

TABLE 5

| | Change in Number of Articles in the Two Top 100 Samples: (six-100) − (ten-100) | | | | | | | |
|---|---|---|---|---|---|---|---|---|
| Journal | 2001 | 2002 | 2003 | 2004 | 2005 | 2006 | Total | Stdev |
| A&A | 1 | -2 | 1 | 1 | 4 | 2 | 7 | 1.9 |
| AJ | -1 | -1 | 1 | 0 | 0 | 1 | 0 | 0.9 |
| ApJ | 0 | -1 | -4 | 0 | -5 | -6 | -16 | 2.7 |
| ApJS | -1 | 1 | 0 | 2 | -1 | 0 | 1 | 1.2 |
| ARA&A | 0 | 2 | 0 | 1 | 4 | 2 | 9 | 1.5 |
| JCosAstPP | 0 | 0 | 1 | 0 | 1 | 1 | 3 | 0.5 |
| MNRAS | 1 | -1 | 0 | -2 | -3 | -3 | -8 | 1.6 |
| Nature | 0 | 0 | -3 | 2 | 0 | 3 | 2 | 2.1 |
| PASP | 1 | 0 | 4 | 1 | 0 | 0 | 6 | 1.5 |
| Science | 0 | 1 | 1 | -7 | 1 | -1 | -5 | 3.1 |

TABLE 6

NUMBER OF AUTHORS AND QUARTILES OF AUTHOR DISTRIBUTIONS FOR
TOP 100 PAPERS FOR EACH YEAR

| Year | Total Authors | Total Unique Authors | Q1 | | Q2 | | Q3 | | Q4 | |
|------|------|------|-------|---------|-------|---------|-------|---------|-------|---------|
| | | | Value | Authors | Value | Authors | Value | Authors | Value | Authors |
| 2009 | 2668 | 1468 | 4.0 | 68 | 8.0 | 138 | 19.0 | 343 | 242 | 2032 |
| 2008 | 2830 | 1898 | 4.8 | 63 | 9.0 | 158 | 19.0 | 369 | 457 | 2236 |
| 2007 | 2826 | 1895 | 5.0 | 74 | 15.0 | 230 | 29.0 | 541 | 450 | 1986 |
| 2006 | 2349 | 1493 | 5.0 | 82 | 10.5 | 198 | 23.3 | 392 | 174 | 1587 |
| 2005 | 1702 | 1177 | 3.8 | 58 | 8.5 | 141 | 24.0 | 362 | 160 | 1199 |
| 2004 | 1598 | 1220 | 4.0 | 65 | 7.0 | 132 | 16.3 | 257 | 160 | 1144 |
| 2003 | 1283 | 889 | 3.8 | 54 | 7.0 | 122 | 16.0 | 265 | 193 | 827 |
| 2002 | 1192 | 726 | 2.0 | 44 | 4.0 | 80 | 13.3 | 183 | 194 | 841 |
| 2001 | 991 | 650 | 3.0 | 48 | 4.0 | 78 | 10.3 | 158 | 69 | 699 |
| 2000 | 798 | 671 | 2.0 | 40 | 4.0 | 75 | 6.0 | 120 | 146 | 564 |
| 1995 | 664 | 601 | 2 | 42 | 3 | 65 | 6.25 | 110 | 40 | 447 |
| 1990 | 315 | 275 | 2 | 29 | 2.5 | 50 | 4 | 82 | 21 | 154 |

TABLE A1

COMPARISON OF CITATIONS BETWEEN ADS AND ISI

| Year | Number of Papers in Common | Number of Citations | | Ratio | σ |
|------|------|------|------|------|------|
| | | ADS | ISI | | |
| 2001 | 19 | 5629 | 4621 | 1.22 | 0.08 |
| 2002 | 15 | 2765 | 2205 | 1.25 | 0.13 |
| 2003 | 25 | 5302 | 4133 | 1.28 | 0.09 |
| 2004 | 20 | 3513 | 2621 | 1.34 | 0.08 |
| 2005 | 29 | 2852 | 1836 | 1.55 | 0.20 |
| 2006 | 24 | 1898 | 873 | 2.17 | 0.48 |

TABLE A2

THE PAPERS RANKED 1ST, 50TH, AND 100TH FOR EACH YEAR FROM 2000 TO 2009

| Year & Rank | Times Cited | Page Length | Authors[§] | Full Reference | Comments |
|---|---|---|---|---|---|
| 2000.001 | 2135 | 9 | 146 | York, et al. AJ, 120, 1579 | SDSS |
| 2000.050 | 259 | 10 | 3 | Norris, et al. ApJ, 534, 248 | GRBs, data from Space |
| 2000.100 | 192 | 20 | 1 | Richards, ApJ, 533, 611 | Extra-galactic, Radio survey |
| 2001.001 | 1541 | 26 | 15 | Freedman, et al. ApJ | $H_0$ Key Project, HST |
| 2001.050 | 253 | 57 | 2 | Reiputh & Bally, ARA&A, 39, 403 | H-H objects, review |
| 2001.100 | 192[†] | 12 | 3 | Vikhlinin, et al. ApJ, 551.160 | Extra-galactic, Chandra |
| 2002.001 | 1059 | 64 | 194 | Stoughton, et al. AJ, 123, 485 | SDSS, survey |
| 2002.050 | 233[†] | 17 | 3 | Klypin, et al. ApJS, 148, 175 | Galactic structure, theory |
| 2002.100 | 171[†] | 35 | 8 | Feigelson, et al. ApJ, 574, 258 | Stellar, ISM, Chandra |
| 2003.001 | 4962 | 20 | 17 | Spergel, et al. ApJS, 148, 175 | WMAP, survey |
| 2003.050 | 262 | 25 | 23 | Kennicutt, et al. PASP, 115, 928 | SINGS, survey |
| 2003.100 | 178 | 7 | 3 | Cohen, et al. AJ, 126, 1607 | 2MASS, survey |
| 2004.001 | 1479 | 23 | 20 | Riess, et al. ApJ, 607, 665 | Supernova, HST |
| 2004.050 | 195 | 21 | 20 | Golimowski, et al. AJ, 127, 3516 | Stellar, IR observations |
| 2004.100 | 144[†] | 39 | 4 | Gibb, et al. ApJS, 151, 35 | Stellar, ISM, ISO |
| 2005.001 | 770 | 15 | 48 | Eisenstein, et al. ApJ, 633, 560 | SDSS, BAO, survey |
| 2005.050 | 171 | 50 | 2 | Beers & Christlieb, ARA&A, 43, 531 | Low metallicity stars |
| 2005.100 | 128 | 3 | 4 | Tsiganis, et al. Nature, 435, 459 | Solar system dynamics, theory |
| 2006.001 | 1002 | 21 | 29 | Skrutskie, et al. AJ, 131, 1163 | 2MASS, survey |
| 2006.050 | 132 | 5 | 12 | Lovis, et al. Nature, 441, 305 | Extra-solar planet discovery |
| 2006.100 | 103[†] | 15 | 5 | Cattaneo, et al. MNRAS, 370, 1561 | Extra-galactic, theory |
| 2007.001 | 2051 | 32 | 20 | Spergel, et al. ApJS, 170, 377 | WMAP, survey |
| 2007.050 | 92 | 26 | 27 | Calzetti, et al. ApJ, 666, 870 | Stellar, Spitzer, HST |
| 2007.100 | 67[†] | 4 | 5 | Johansen, et al. Nature | Proto-planets, theory |
| 2008.001 | 314 | 17 | 170 | Adelman-McCarthy, et al. ApJS, 175, | SDSS, survey |
| 2008.050 | 47[†] | 10 | 102 | Frieman, et al. AJ, 135, 338 | SDSS, supernovae, survey |
| 2008.100 | 36[†] | 25 | 5 | Liang, et al. ApJ, 675, 528 | GRBs, Swift |
| 2009.001 | 633 | 47 | 19 | Komatsu, et al. ApJS, 180, 330 | WMAP, survey |
| 2009.050 | 18[†] | 14 | 41 | Salvato, et al. ApJ, 690, 1250 | Extra-galactic, XMM, survey |
| 2009.100 | 13[†] | 5 | 8 | Mumma, et al. Science, 323, 104 | Mars, ground based telescopes |

[§] For papers with more than about 10 authors the numbers in this column may be inaccurate by a couple of percent.

[†] There is at least one other paper amongst the top 100 for this year with this many citations.

# FIGURE CAPTIONS

**Figure 1** – The full bar height and the left hand scale indicate the total number of authors' names of the 100 most cited articles in astronomy and astrophysics for each year. The cross hatched area indicates how many of these names are unique for the year. The right hand scale and the symbols give membership numbers for the AAS and the IAU and the total number of articles published each year by the journals listed in Table 1; these are in the subset of astronomy journals that were searched for the top 100.

**Figure 2** – For each two year period indicated in the figure legend, the cumulative sum of the average number of papers in the top 100 sample with less than or equal to "X" number of authors is plotted. The dashed line at 100 is shown for convenience. With the exception of the two most recent years, this figure clearly illustrates the systematic decline amongst the top 100 papers in small numbers of authors with a concurrent increase in papers with large numbers of authors.

**Figure 3** – This figure illustrates the yearly change in number of papers with "X" number of authors for X between 1 and 30 for papers in the top 100 sample. This quantity is just the slope of the regression line for each value of X. The doubling of the time baseline by the addition of data for 1990 and 1995 considerably reduces the scatter for X=6 and greater. In either case, though, the inferences to be drawn are the same: a steady decline with time for papers with 4 and fewer authors accompanied by small yearly increases in the number of papers with more than 6 or 7 authors.

**Figure 4** – The number of authors for each of the top 100 papers for a year has been normalized by dividing by the total number of authors for that year. Then the quartiles for the distributions of normalized authors are calculated and plotted. Quartiles are defined such that for Q1 25% of the papers in that year have Q1 or few authors, etc. The Q2 values are the medians. The total number of authors for each year has been normalized by the total number for all 10 years. Note the close similarity of the shapes of the normalized distributions.

**Figure 5** – The number of citations to each of the top 100 papers for each year has been normalized by the sum of all citations to the 100 papers. For convenience these normalized citation counts have all been multiplied by 1000. The vertical axis is logarithmic. For ranks between about 35 and 100 the year-to-year scatter is comparable in size to the symbols.

**Figure 6** – The same as Figure 5 except for just the top 30 papers. The sold red line is average of the normalized citation numbers for each rank over the 10 year period. The black line is a power-law fit to the data for ranks 6 to 100 extended (dashed line) to the first rank. It and the average line are indistinguishable for ranks 6 and higher.

**Figure 7** – The red line is the average of the data points as in Figure 6. The error bars show the standard deviation of the 10 points at each rank. This figure shows that the scatter in the normalized citation counts is significant only for the most highly cited papers. Thus although the total number of citations to the top 100 papers increases strongly with time, the shape of the distributions over rank are essentially identical except for the most cited papers.

**Figure 8** – This is the same as Figure 6, but now only the top 8 papers are shown. Note that the points with the largest scatter all lie above the average line and well above the extrapolation of the power law fit.

**Figure 9** – For the ten papers at each citation rank the points are the average page length of the ten papers at each citation rank. It was not necessary to normalize the page lengths from different years as no significant time dependence was found for this quantity. The solid red line is the regression fit; the two dashed lines indicate the 95% confidence limits for the fit.

**Figure 10** – The average of the normalized number of authors for the 10 papers at each citation rank is plotted. The red line is a power law fit to the data.

**Figure 11** – The list of the averages of the normalized number of authors for the 10 papers at each citation rank (see previous figure) is rank ordered such that a rank of "1" is the greatest number of authors. This is then plotted against the citation rank. The solid circles are for 2000 to 2009. The small x's are the two year averages for 1990 and 1995. In order to emphasize the correlation between the papers with the most citations and the most authors for the 2000 to 2009 sample the dashed lines have been drawn.

**Figure 12** – This figure illustrates the yearly growth in the number of articles published in two samples: All of those which JCR classifies as astronomy and astrophysics and the sample from which the top 100 articles are drawn. The latter sample has about one-third as many journals as the first sample but about three-quarters as many articles. The lines are the linear regression fits to the two data sets. They have statistically identical slopes.

**Figure 13** – This figure illustrates the growth in usage of the ADS web site in three ways: The number of users, the number of abstracts they viewed, and the number of articles for which they downloaded the full text. The leveling off in the number of users over the past few years mimics the leveling off in the number of author for the top 100 papers in the same time period.

FIGURE 1

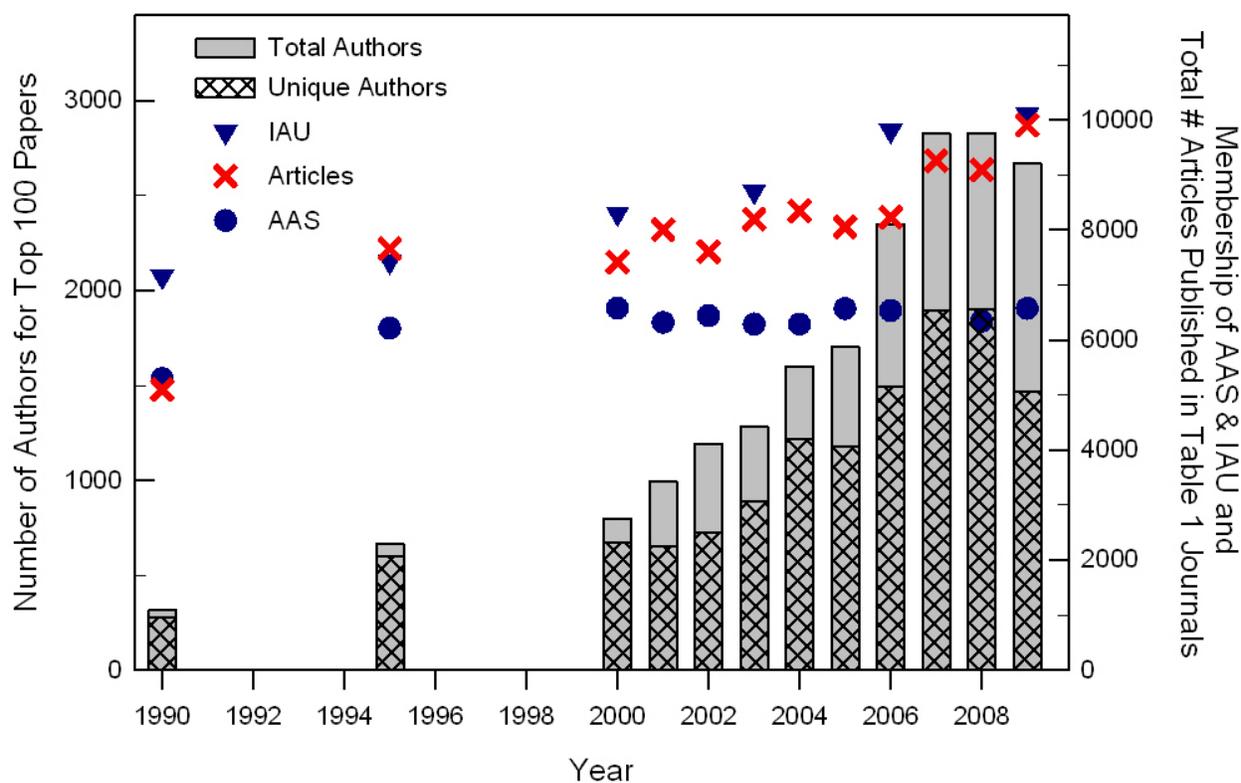

**Figure 1** – The full bar height and the left hand scale indicate the total number of authors' names of the 100 most cited articles in astronomy and astrophysics for each year. The cross hatched area indicates how many of these names are unique for the year. The right hand scale and the symbols give membership numbers for the AAS and the IAU and the total number of articles published each year by the journals listed in Table 1; these are in the subset of astronomy journals that were searched for the top 100.

FIGURE 2

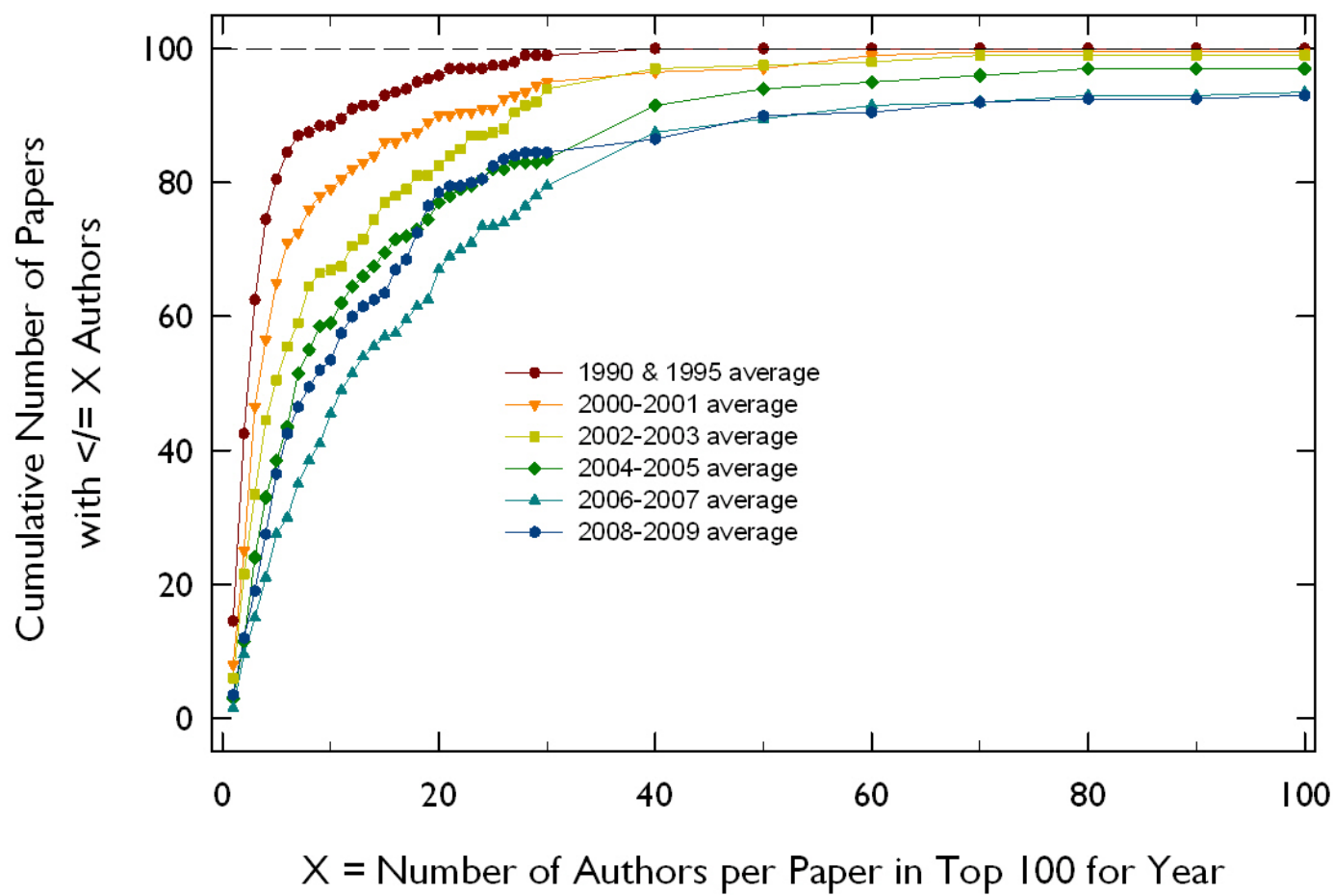

**Figure 2** – For each two year period indicated in the figure legend, the cumulative sum of the average number of papers in the top 100 sample with less than or equal to "X" number of authors is plotted. The dashed line at 100 is shown for convenience. With the exception of the two most recent years, this figure clearly illustrates the systematic decline amongst the top 100 papers in small numbers of authors with a concurrent increase in papers with large numbers of authors.

FIGURE 3

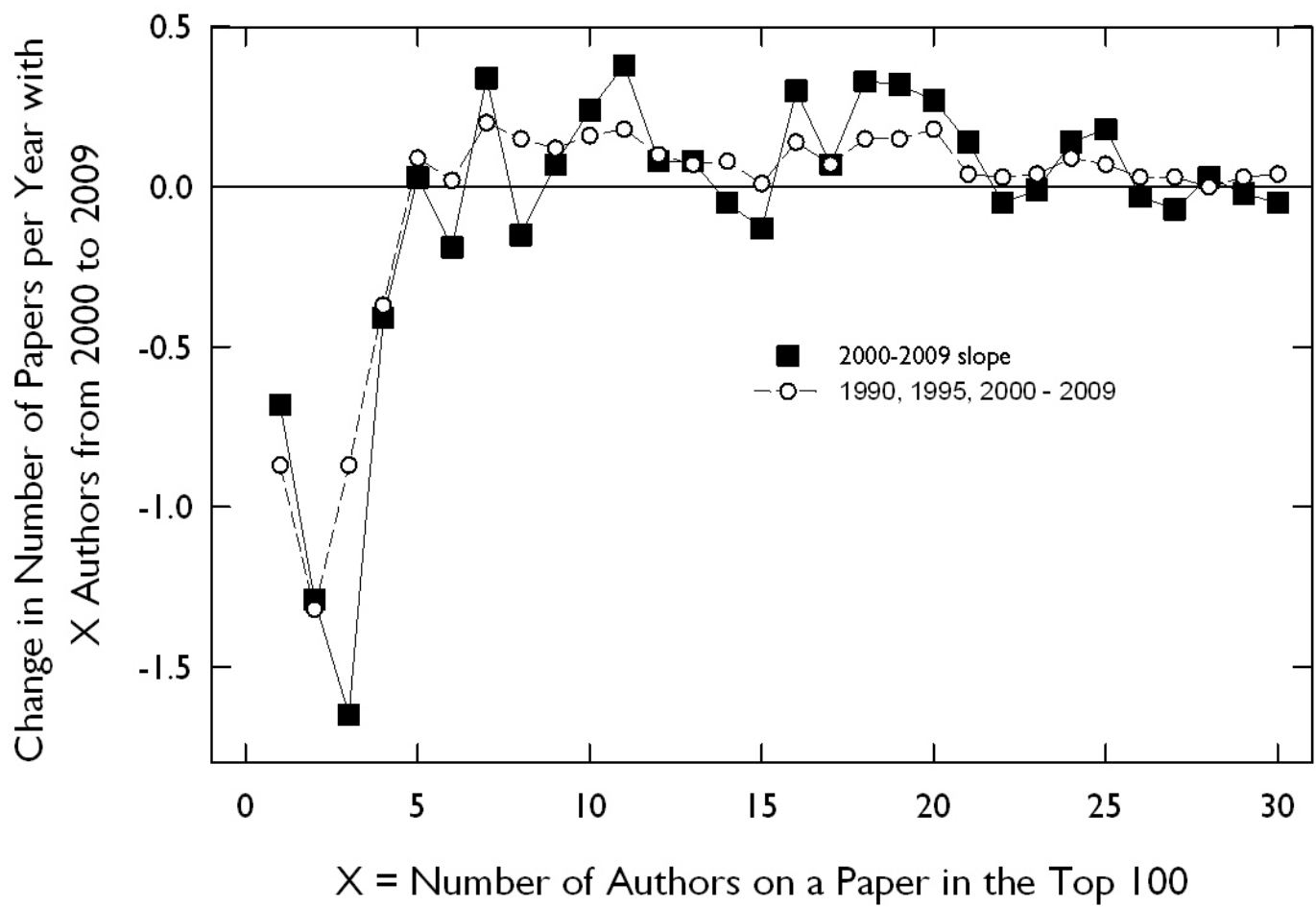

**Figure 3** – This figure illustrates the yearly change in number of papers with "X" number of authors for X between 1 and 30 for papers in the top 100 sample. This quantity is just the slope of the regression line for each value of X. The doubling of the time baseline by the addition of data for 1990 and 1995 considerably reduces the scatter for X=6 and greater. In either case, though, the inferences to be drawn are the same: a steady decline with time for papers with 4 and fewer authors accompanied by small yearly increases in the number of papers with more than 6 or 7 authors.

FIGURE 4

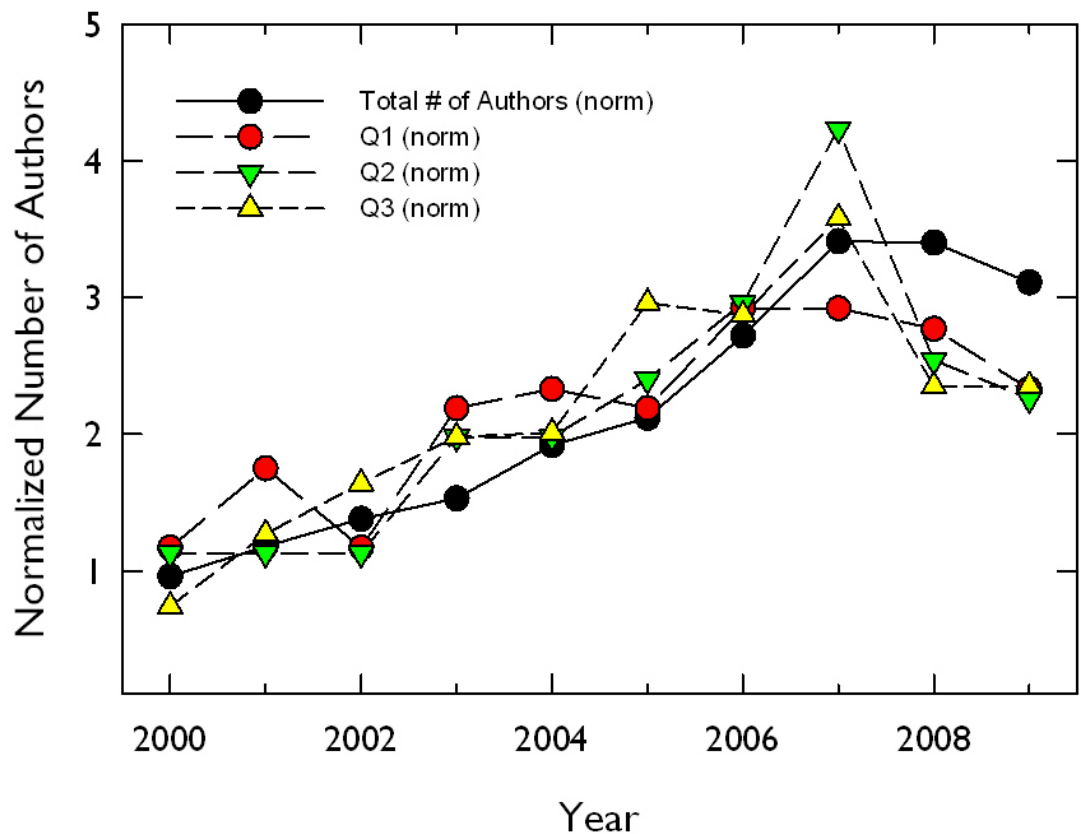

**Figure 4** – The number of authors for each of the top 100 papers for a year has been normalized by dividing by the total number of authors for that year. Then the quartiles for the distributions of normalized authors are calculated and plotted. Quartiles are defined such that for Q1 25% of the papers in that year have Q1 or few authors, etc. The Q2 values are the medians. The total number of authors for each year has been normalized by the total number for all 10 years. Note the close similarity of the shapes of the normalized distributions.

FIGURE 5

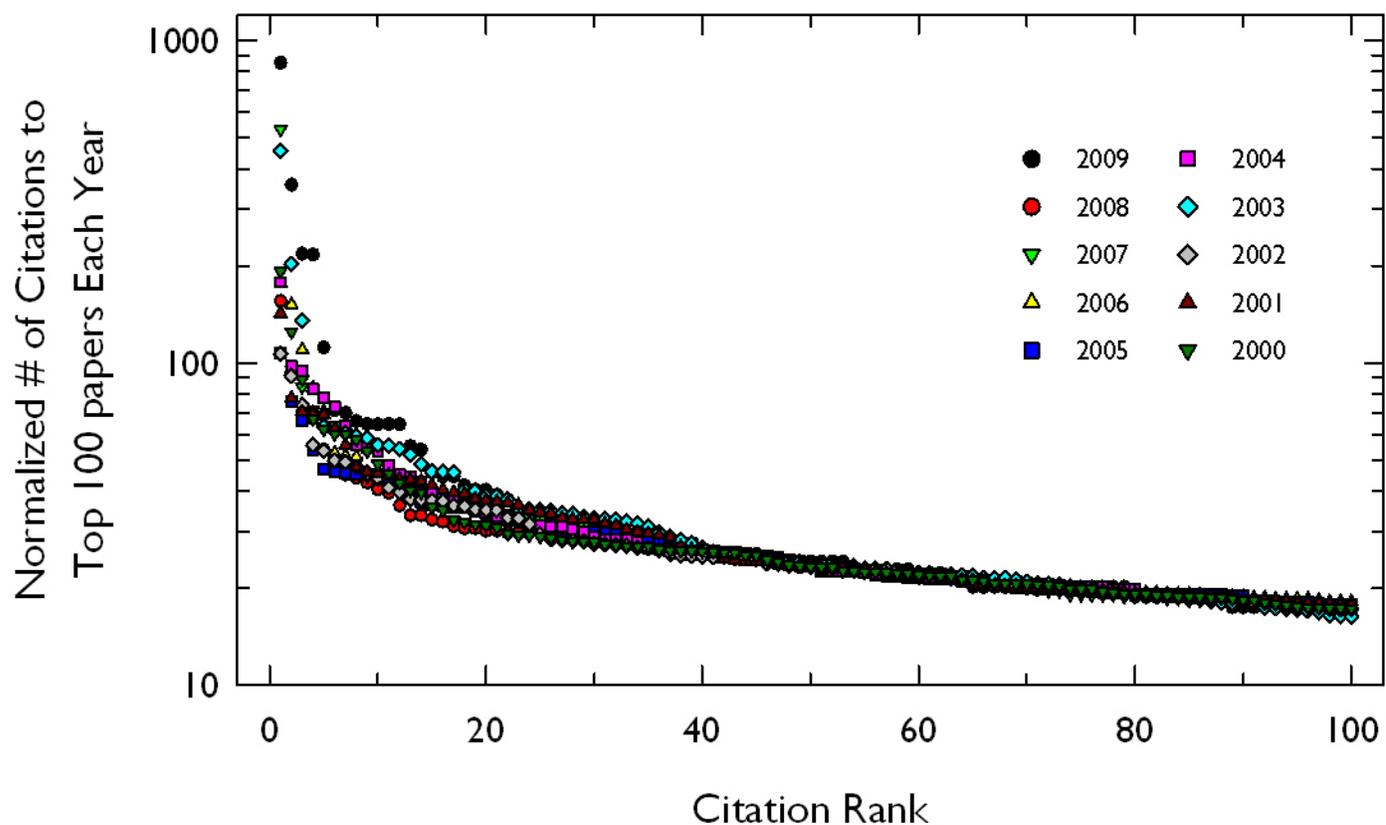

**Figure 5** – The number of citations to each of the top 100 papers for each year has been normalized by the sum of all citations to the 100 papers. For convenience these normalized citation counts have all been multiplied by 1000. The vertical axis is logarithmic. For ranks between about 35 and 100 the year-to-year scatter is comparable in size to the symbols.

FIGURE 6

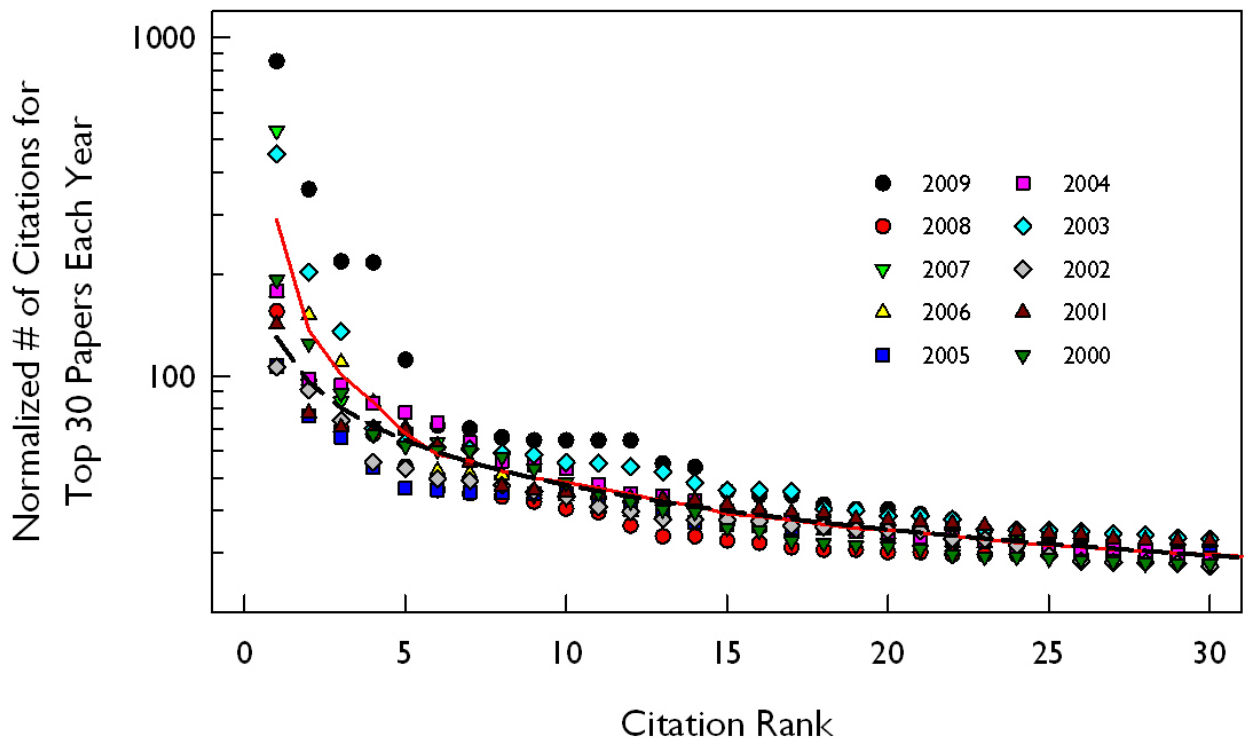

**Figure 6** – The same as Figure 5 except for just the top 30 papers. The sold red line is average of the normalized citation numbers for each rank over the 10 year period. The black line is a power-law fit to the data for ranks 6 to 100 extended (dashed line) to the first rank. It and the average line are indistinguishable for ranks 6 and higher.

FIGURE 7

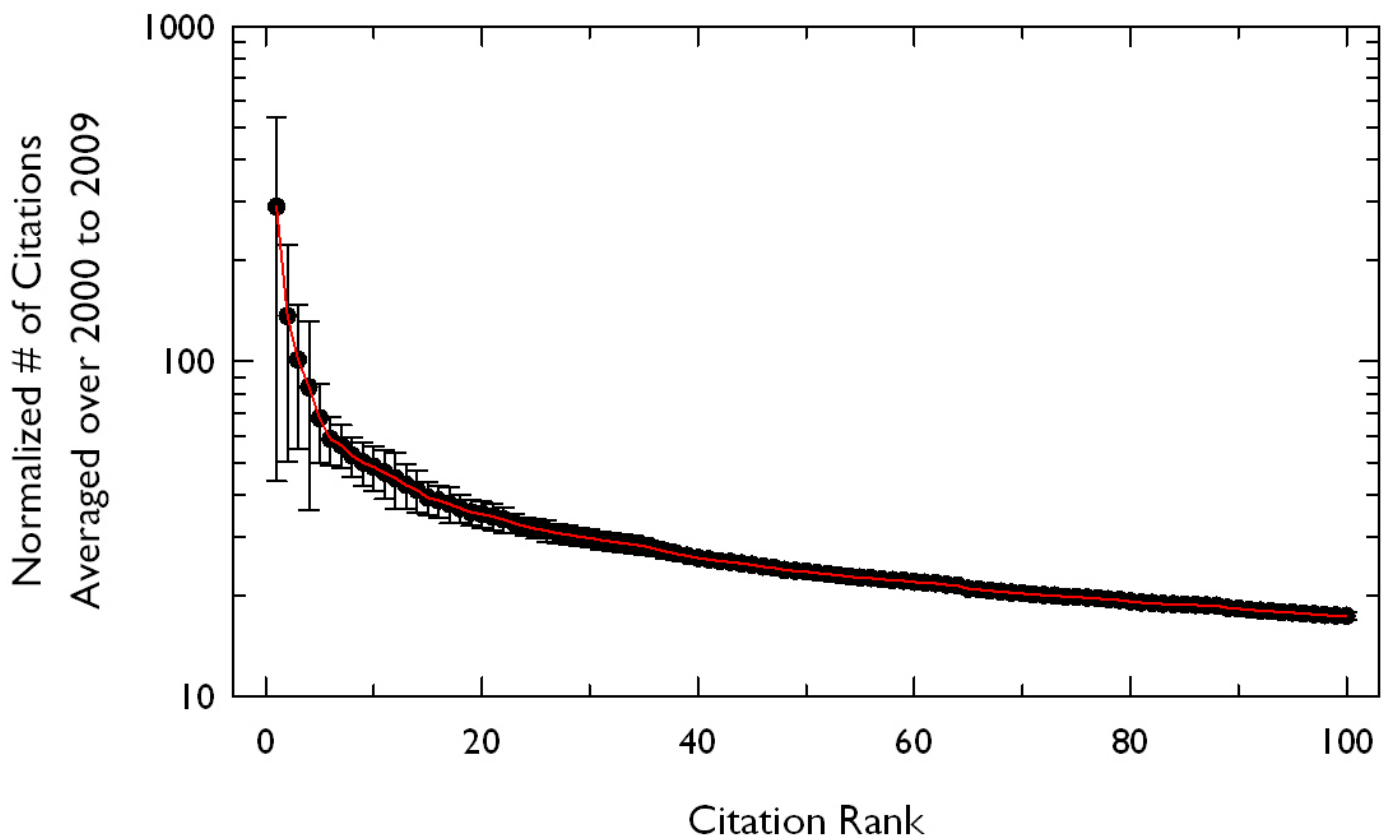

**Figure 7** – The red line is the average of the data points as in Figure 6. The error bars show the standard deviation of the 10 points at each rank. This figure shows that the scatter in the normalized citation counts is significant only for the most highly cited papers. Thus although the total number of citations to the top 100 papers increases strongly with time, the shape of the distributions over rank are essentially identical except for the most cited papers.

FIGURE 8

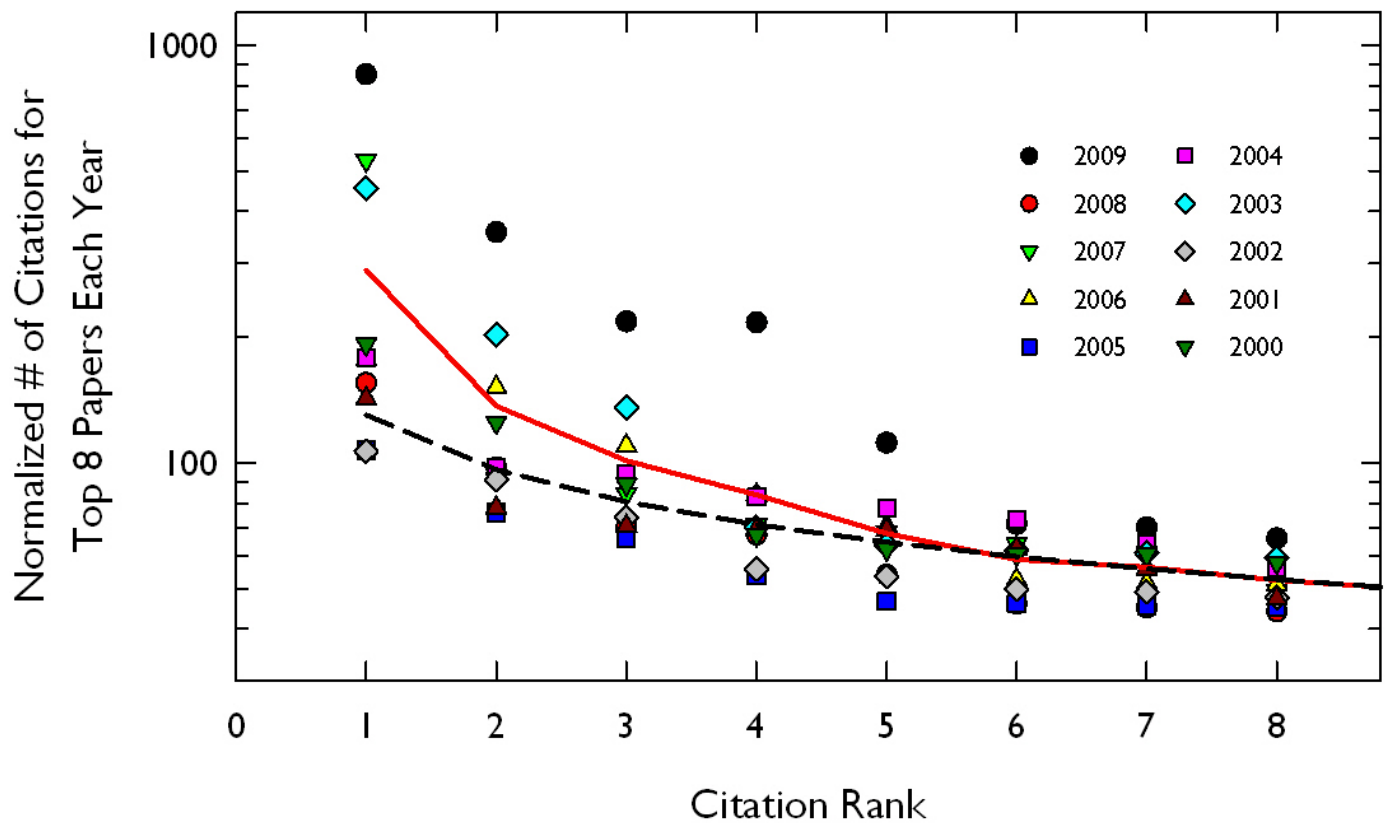

**Figure 8** – This is the same as Figure 6, but now only the top 8 papers are shown. Note that the points with the largest scatter all lie above the average line and well above the extrapolation of the power law fit.

FIGURE 9

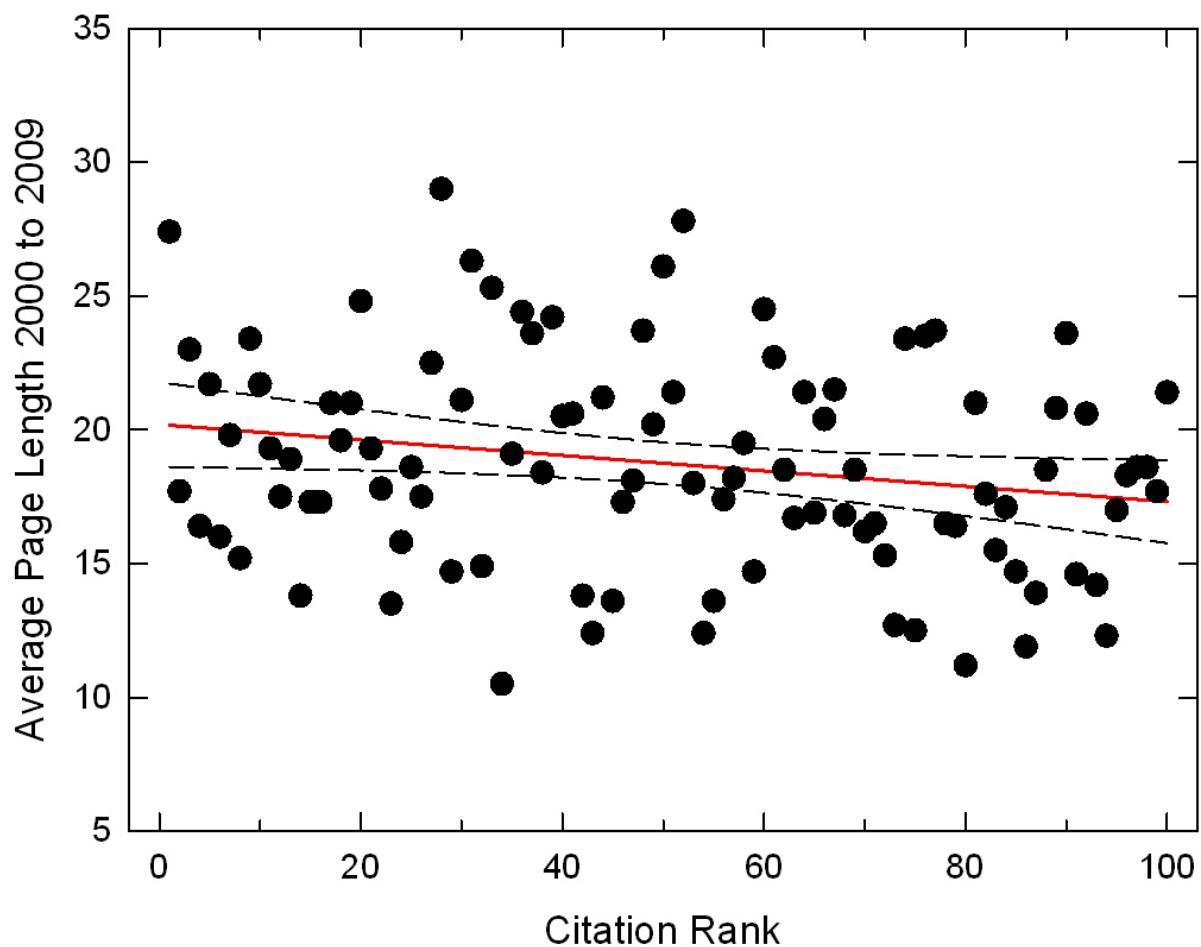

**Figure 9** – For the ten papers at each citation rank the points are the average page length of the ten papers at each citation rank. It was not necessary to normalize the page lengths from different years as no significant time dependence was found for this quantity. The solid red line is the regression fit; the two dashed lines indicate the 95% confidence limits for the fit.

FIGURE 10

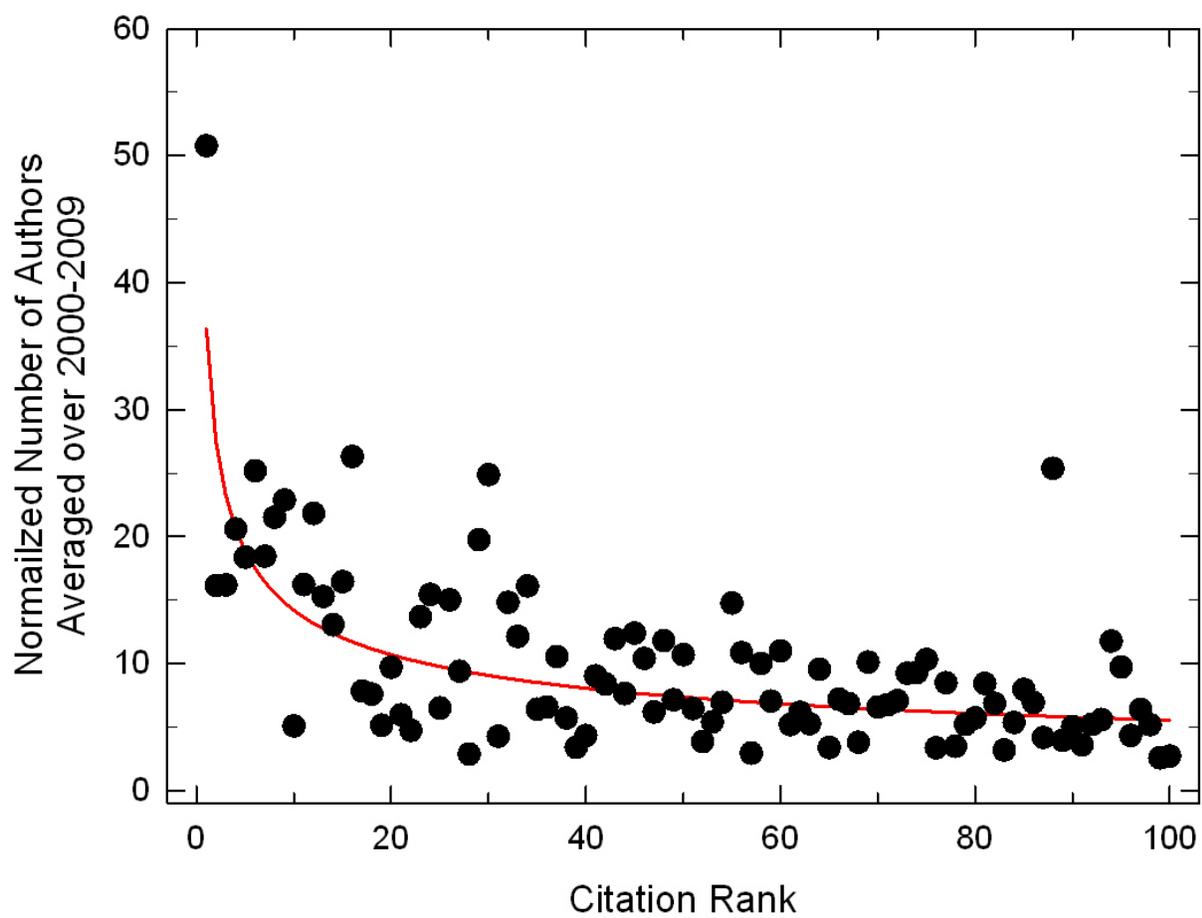

**Figure 10** – The average of the normalized number of authors for the 10 papers at each citation rank is plotted. The red line is a power law fit to the data.

FIGURE 11

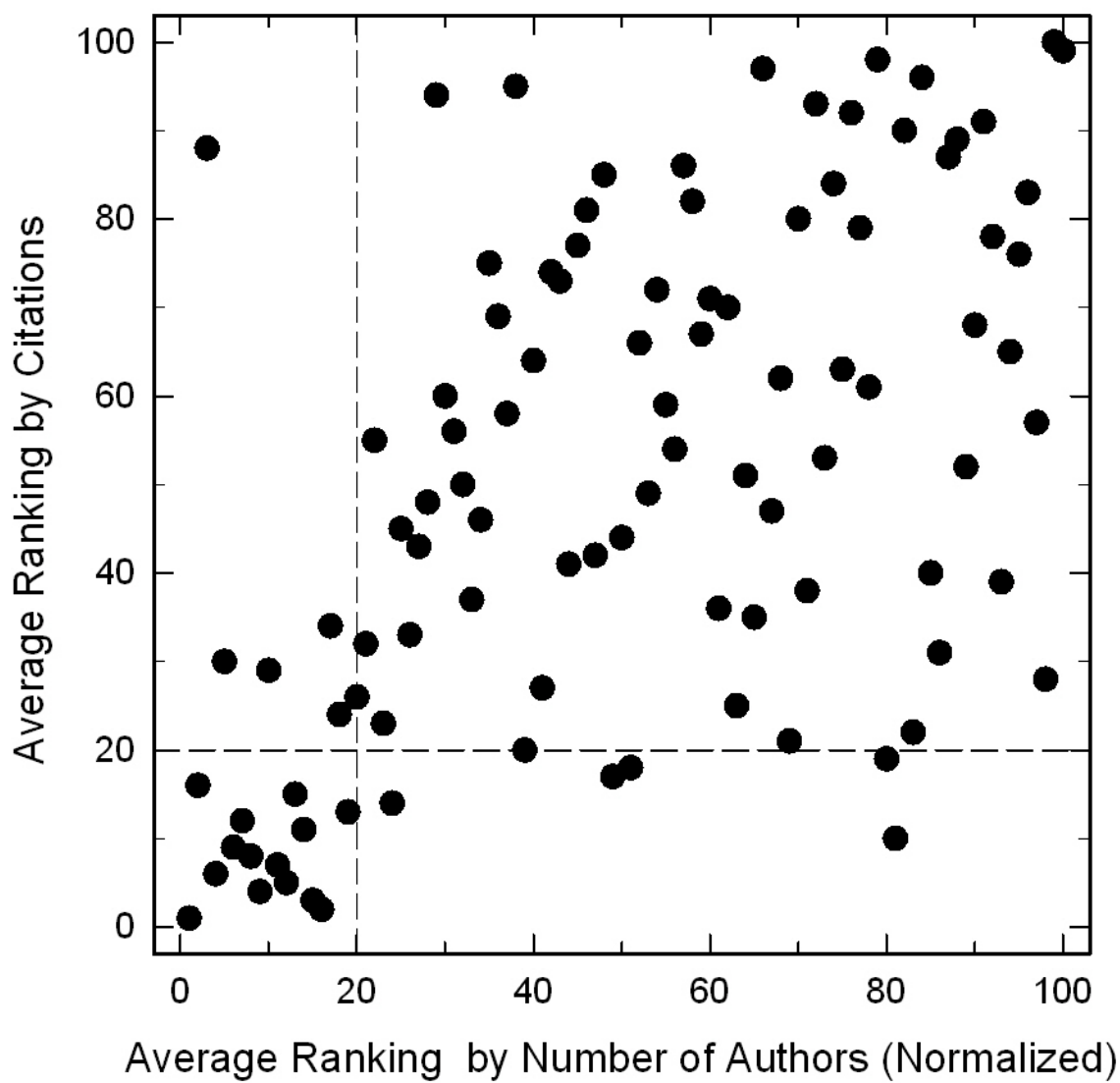

**Figure 11** – The list of the averages of the normalized number of authors for the 10 papers at each citation rank for 2000 to 2009 (see previous figure) is rank ordered such that a rank of "1" is the greatest number of authors. This is then plotted against the citation rank. To emphasize the correlation between the papers with the most citations and the most authors for the 2000 to 2009 sample the dashed lines have been drawn.

FIGURE 12

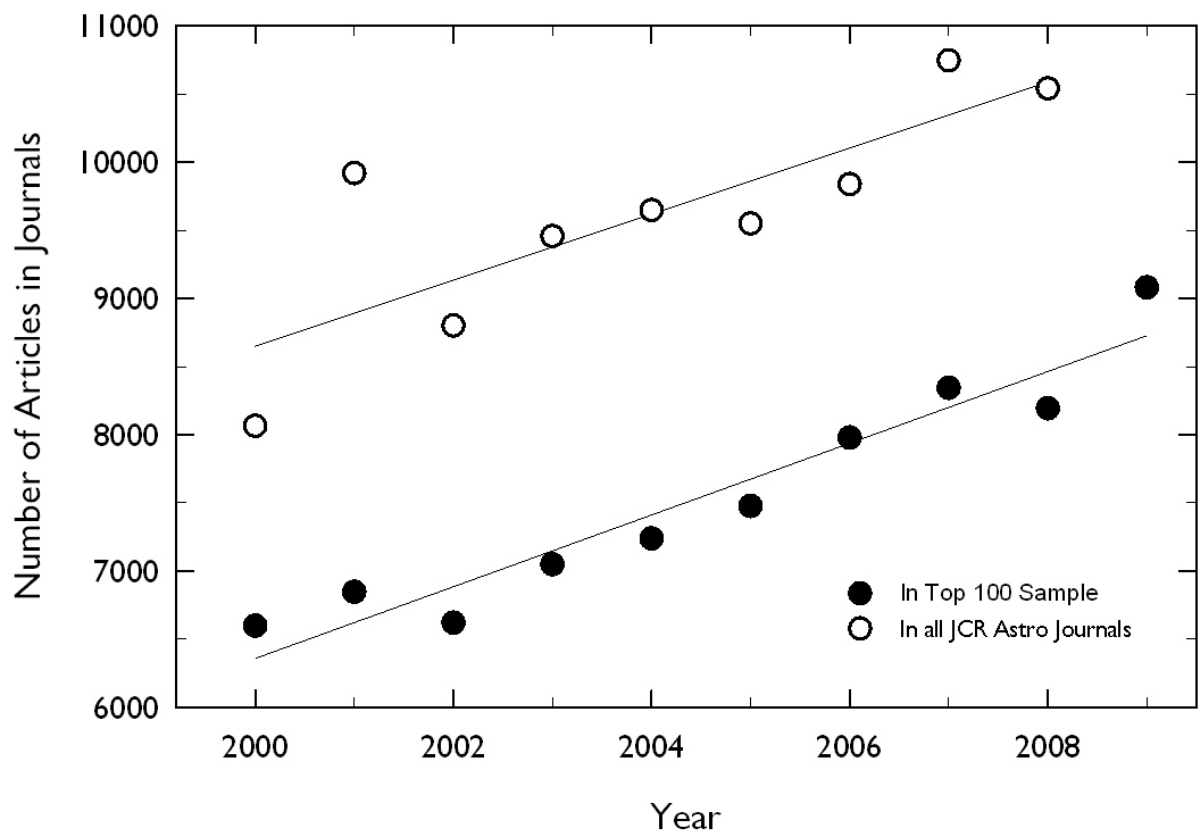

**Figure 12** – This figure illustrates the yearly growth in the number of articles published in two samples: All of those which JCR classifies as astronomy and astrophysics and the sample from which the top 100 articles are drawn. The latter sample has about one-third as many journals as the first sample but about three-quarters as many articles. The lines are the linear regression fits to the two data sets. They have statistically identical slopes.

FIGURE 13

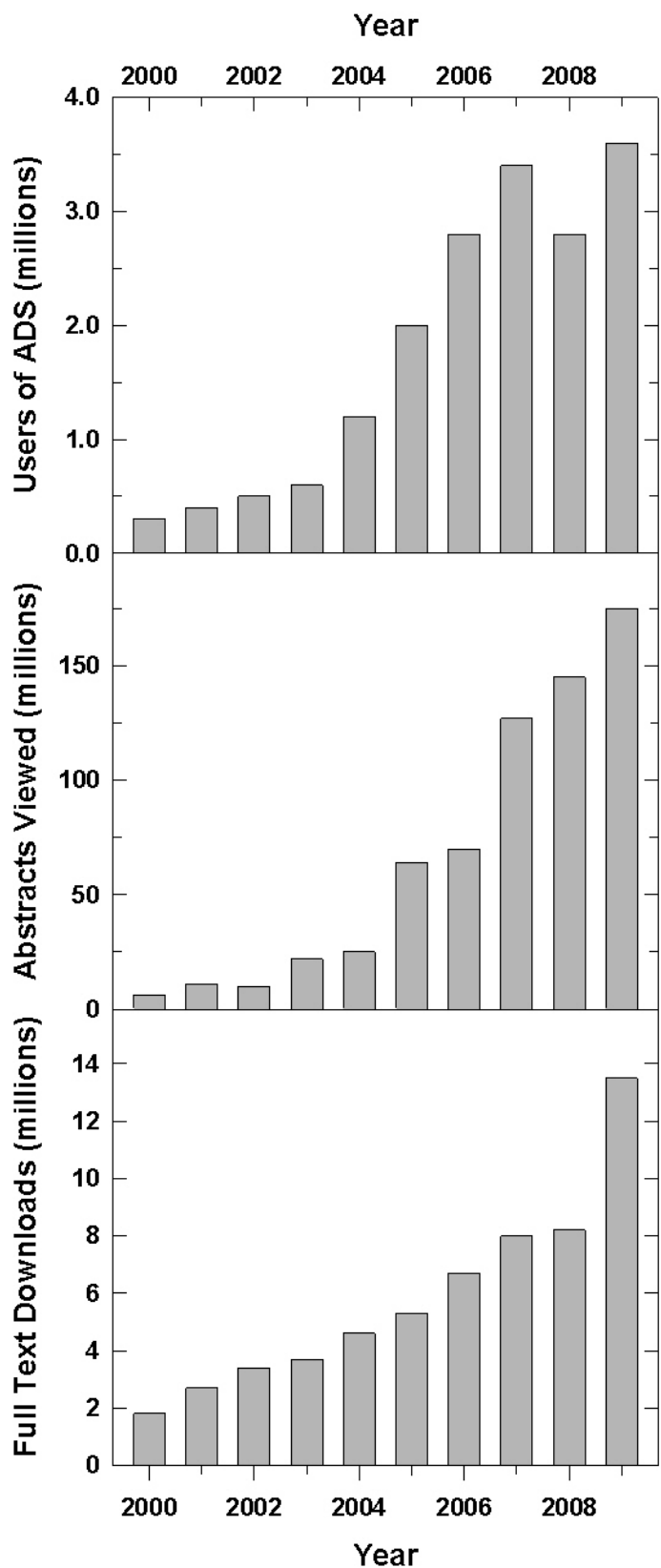

**Figure 13** – This figure illustrates the growth in usage of the ADS web site in three ways: The number of users, the number of abstracts they viewed, and the number of articles for which they downloaded the full text. The leveling off in the number of users over the past few years mimics the leveling off in the number of author for the top 100 papers in the same time period.